\documentclass[amsmath,11pt,amssymb,preprint,nofootinbib,,floatfix]{revtex4}
\usepackage{amsfonts}
\usepackage{lipsum}
\allowdisplaybreaks
\usepackage{graphicx} 
\usepackage{epsfig}
\usepackage{amsmath}
\usepackage{bm}
\usepackage[symbol]{footmisc}
\usepackage{xcolor}

\begin{document}

\title{Open Bottom Mesons and Upsilon States in Hot Magnetized Strange Hadronic Matter}
\author{Amal Jahan C.S.}
\email{amaljahan@gmail.com}
\author{Amruta Mishra}
\email{amruta@physics.iitd.ac.in}
\affiliation{Department of Physics, Indian Institute of Technology,Delhi, Hauz Khas, New Delhi - 110016, India}

\begin{abstract}

The masses of open bottom mesons ($B$($B^+$,$B^0$), $\bar{B}$($B^-$,$\bar{B^0}$), $B_s$(${B_{s}}^0$, $\bar{{B_s}^0}$)) and upsilon states ($\Upsilon(1S)$, $\Upsilon(2S)$, $\Upsilon(3S)$, $\Upsilon(4S)$, and $\Upsilon(1D)$) are investigated in the isospin asymmetric strange hadronic medium at finite temperature in the presence of strong magnetic fields using a chiral effective Lagrangian approach. Here the chiral $SU(3)$ Lagrangian is generalized to include the bottom sector to incorporate the interactions of the open bottom mesons with the magnetized medium. At finite temperature, the number density and scalar density of baryons are expressed in terms of thermal distribution functions. For charged baryons, the magnetic field introduces contribution from Landau energy levels. The masses of the open bottom mesons get modified through their interactions with the baryons and the scalar mesons, which undergo modifications in a magnetized medium. The charged $B^+$, $B^-$ mesons have additional positive mass shifts due to Landau quantization in the presence of the magnetic field. The medium mass shift of the upsilon states originates from the modification of the gluon condensates simulated by the variation of dilaton field ($\chi$) and a quark mass term in the magnetized medium. The open bottom mesons and upsilon states experience a mass drop in the magnetized medium. The masses of these mesons initially increase with a rise in temperature, and beyond a high value of temperature, their masses are observed to drop. When the temperature is below 90 MeV, the in-medium masses of the mesons increase with an increase in the magnetic field. However, at high temperatures (T $>$ 90 MeV), the masses are observed to drop with an increase in the magnetic field. These in-medium effects can have observable effects in asymmetric heavy-ion collisions planned at compressed baryonic matter experiments at FAIR at the future facility of GSI.\\\\

\end{abstract}
\maketitle

\section{INTRODUCTION}

The influence of strong magnetic fields on the properties of QCD matter is relevant in non-central heavy-ion collision experiments. The magnitude of the magnetic fields could reach up to $eB \sim 2{m_{\pi}}^2$ $\sim 6\times 10^{18}$ Gauss in the Relativistic Heavy Ion Collider(RHIC) at Brookhaven National Laboratory (BNL) and eB $\sim 15{m_{\pi}}^2$ $\sim 10^{19}$  Gauss in the Large Hadron Collider(LHC) at CERN \cite{kharzeev,skokov}. In these studies, the strength of the magnetic field depends upon the collision energy as well as the value of the impact parameter. The event-averaged magnetic field is produced perpendicular to the reaction plane, and the dominant contribution to the magnetic field arises from the spectator particles, which do not undergo collision. Another physical system where intense magnetic fields are of significant importance is the magnetars, where the strength of magnetic fields may reach $10^{15}-10^{16}$ Gauss at the surface \cite{magnetar1,magnetar2}. The effect of strong magnetic fields on the equation of state (EOS) of the dense hadronic matter in neutron stars is investigated in Ref \cite{Shapiro_Abrahams, Wei, Mao, Mao2, Broderick2,Broderick1, Lopes, Gomes}. The effects of magnetic fields on the EOS of hybrid stars are studied in Ref \cite{Dexheimer}. 

 The phase structure of the QCD vacuum in the magnetic field at low temperature is studied using chiral perturbation theory in Ref \cite{Agasian}. The effects of the magnetic field on the QCD phase diagram and phase transition at finite temperature are studied using a linear sigma model coupled to quarks and Polyakov loops in \cite{Mizher} as well as in lattice QCD \cite{Balimag}. The role of the magnetic field in asymptotic freedom is investigated in Ref \cite{Andreichikov}. Since charm and bottom quarks are produced at the early stages of the heavy-ion collision where the strength of the magnetic field is sufficiently high, the heavy flavor mesons may experience the effect of magnetic fields \cite{Cho}. The chiral condensates in QCD, which acts as an order parameter for the amount of spontaneous chiral symmetry breaking, is altered through magnetic catalysis and inverse magnetic catalysis in the presence of magnetic fields \cite {Balimag2, GusyninNJL, Holographic}. The gluon condensates, which play a role in scale symmetry breaking, are also modified by strong magnetic fields \cite{Balimag, gluoncatalysis2}. As the hadrons produced in these heavy-ion collision experiments interact with these condensates, their properties also modify in the presence of the magnetic field at finite baryon density and temperature. 

The effect of external magnetic fields on the mass of charged $B$ mesons was studied using the QCD sum rules (QSR) approach \cite{Machado}. Here the modification of the quark propagator on the Operator Product Expansion side and the modification of the meson propagator on the phenomenological side were taken into account. The modification of chiral condensates in the presence of the magnetic field was also considered in \cite{Machado}. The properties of open charm mesons and heavy quarkonia have also been investigated in the QCD Sum rule approach, including the effects of the magnetic field \cite{Gubler, Cho, AM_sumrule_spinmixing, bottomonia_sumrule_Pallabi, Rajeshkumar}. In Ref \cite{Gubler, Cho, AM_sumrule_spinmixing,bottomonia_sumrule_Pallabi}, the magnetically induced mixing effects between the pseudoscalar and the vector mesons were considered. The contribution of the Landau levels was also taken into account to study the mass spectra of charged $D$ mesons in Ref \cite{Gubler}. In Ref \cite{AM_sumrule_spinmixing,bottomonia_sumrule_Pallabi, Rajeshkumar, Rajeshkumar2}, the masses of heavy quarkonia in the magnetized nuclear medium are considered using the QSR approach.

The properties of charmonium and bottomonium states have been investigated using potential models in magnetic fields \cite{Alford, Bonati, Suzuki, Yoshida, Machado2}. The Hamiltonian of the heavy quarkonia under an external magnetic field given in Ref \cite{Alford} contains the kinetic energy terms of the meson, the Cornell potential term, the spin-spin interaction term, and the spin-magnetic field interaction term through the magnetic moment of the heavy quarks. The latter term results in the mixing between the spin-singlet and triplet states of the heavy quarkonia. As the value of magnetic moments of the bottom quark is small, bottomonium states experience a smaller mass shift due to the spin-mixing effect than charmonia. In Ref \cite{Machado2}, the effect of the magnetic field on the mass spectrum of pseudoscalar and vector open heavy flavor mesons was considered. The additional effect of magnetic catalysis for the same investigation is included in Ref \cite{Yoshida}. The modification of the Cornell potential in the magnetic field is considered in Ref \cite{Bonati2}. 

 The in-medium properties of open charm mesons, open bottom mesons, and heavy quarkonia have been studied in the chiral effective model  based on a non-linear realization of chiral symmetry and the broken scale invariance of QCD without incorporating the effects of magnetic fields \cite{AM_PRC79_2009,AKumar_AM_EurPhys2011,Akumar_AM_PRC81_2010,Divakar_AM_IntJMod_2014,AM_Divakar_PRC_2014,AM_Divakar_PRC_2015,Divakar_AM_AdvHighEner_2015}. These properties  has been investigated in the cold nuclear matter in the presence of magnetic fields \cite{SReddy,Dhale,Amal1,AM_charmdecaywidths_mag,Amal2_upsilon,AM_SPM3,AM_SPM5,AM_SPM6}. In the chiral effective model, the mass modifications of open charm and open bottom mesons in the hadronic medium arise due to their interaction with the scalar fields and the baryons. The mass modifications of charmonium and upsilon states are due to medium modifications of the dilaton field $\chi$, which simulates the gluon condensates of QCD within the chiral effective model. The magnetic field distinguishes charged baryons from the neutral baryons, with the former having contributions from Landau energy levels. The effects of isospin asymmetry and anomalous magnetic moments of nucleons on the mesons have also been investigated and are significant at large densities. In Ref \cite{AM_SPM4, AM_SPM5}, the effect of magnetically induced spin mixing of pseudoscalar and vector charmonium states in the magnetized nuclear matter and that of open charm mesons in the vacuum was taken into account through a phenomenological interaction Lagrangian. In these studies, the mass shift due to spin mixing is determined using a mixing coupling parameter calculated from the observed value of the radiative decay width of the vector meson decaying to the corresponding pseudoscalar meson. In Ref \cite{Magstrange}, we have investigated the in-medium masses of open charm mesons and charmonia in the cold magnetized strange hadronic matter. 

 In this paper,  we study the mass modifications of the open bottom mesons and upsilon states in the magnetized strange matter at finite temperature. The outline of the paper is as follows: In section II, we describe the chiral effective hadronic model in the presence of the magnetic field with the effect of finite temperature introduced through Fermi distribution functions in the expressions of scalar and number densities of baryons. In section III, we describe the Lagrangian interaction density of the open bottom mesons in the hot magnetized strange hadronic matter. We also describe the dispersion relations and express the self energies of these mesons through which we arrive at their mass modifications. In section IV, we present the leading order in-medium mass shift formula of the upsilon states due to the modification of gluon condensates in the chiral effective model. In section V, we present our discussion and analysis of the results obtained and summarize our findings in section VI.

 \section{THE EFFECTIVE HADRONIC CHIRAL MODEL}
The effective hadronic Lagrangian density in the chiral effective model \cite{SReddy} is given as
\begin{eqnarray}
\mathcal{L_{\textrm{eff}}} = \mathcal{L_\textrm{kin}} + \sum_{W=X,Y,A,V,u}{\mathcal{L_\textrm{BW}}} + \mathcal{L_\textrm{vec}} + \mathcal{L_\textrm{0}}+ \mathcal{L_\textrm{scale break}} + \mathcal{L_\textrm{SB}} + \mathcal{L_\textrm{mag}}.
\label{genlag}
\end{eqnarray}

In this equation, $\mathcal{L_\textrm{kin}}$ refers to the kinetic energy terms of the mesons and baryons. $\mathcal{L_\textrm{BW}}$ is the baryon-meson interaction term, where the index $\mathcal{\textrm{W}}$ covers both spin-0 and spin-1 mesons. Here the baryon masses are generated dynamically through the baryon-scalar meson interactions. $\mathcal{L_\textrm{vec}}$ concerns the dynamical mass generation of the vector mesons through couplings with scalar mesons, apart from bearing the quartic self-interaction terms of these mesons. $\mathcal{L_\textrm{0}}$ contains the meson-meson interaction terms that introduce the spontaneous breaking of chiral symmetry, and $\mathcal{L_\textrm{scale break}}$ incorporates the scale invariance breaking of QCD through a logarithmic potential given in terms of scalar dilaton field $\chi$. $\mathcal{L_\textrm{SB}}$ corresponds to the explicit chiral symmetry breaking term, and $\mathcal{L_\textrm{mag}}$ is the contribution by the magnetic field given as
\begin{eqnarray}
\mathcal{L_\textrm{mag}} =-\bar{\psi_{i}}q_{i}\gamma_{\mu}A^{\mu}\psi_{i}-\frac{1}{4}\kappa_{i}\mu_{N}\bar{\psi_{i}}\sigma^{\mu\nu}F_{\mu\nu}\psi_{i}-\frac{1}{4}F^{\mu\nu}F_{\mu\nu}.
\label{Lmag}
\end{eqnarray}
Here the index i runs over the eight lightest baryons $p$, $n$, $\Lambda$, $\Sigma^{-}$,$\Sigma^{0}$, $\Sigma^{+}$, $\Xi^{-}$, $\Xi^{0}$. The second term in eq.(\ref{Lmag}), which is a tensorial interaction term, is related to the anomalous magnetic moment(AMM) of the baryons. In this term,  $\mu_{N}$ is the nuclear Bohr magneton, given as $\mu_{N}$  = $e/(2m_N)$, where $m_N$ is the vacuum mass of the nucleon. Here $\kappa_{i}$ is the gyromagnetic ratio corresponding to the AMM of the baryons and their values used in our calculations were taken from Ref \cite{Wei,Broderick1}. We choose the magnetic field to be uniform and along the z-axis. We take the vector potential to be $A^{\mu}$ = (0, 0, Bx, 0).

We use the mean-field approximation to simplify the hadronic Lagrangian density under which all the meson fields are considered as classical fields. From the mean-field Lagrangian density, the coupled equations of motion for the scalar fields ( non- strange field $\sigma$, strange field $\zeta$, isovector field $\delta$, dilaton field $\chi$), and vector meson fields ($\omega, \rho, \phi$) are obtained. The equations of motion for the scalar fields $\sigma$, $\zeta$, and $\delta$ are in terms of scalar densities of baryons \cite{Papazoglou,AM_PRC69_2004,Zschiesche}. The equations of motion for vector meson fields $\omega, \rho, \phi$ are in terms of the number densities of baryons \cite{Papazoglou,AM_PRC69_2004,Zschiesche}. The magnetic field introduces Landau quantization in the expressions for number density and scalar density of charged baryons ( i= p, $\Sigma^{-}$, $\Sigma^{+}$, $\Xi^{-}$ ), and at finite temperature, they are given as \cite {Broderick2}
\begin{equation}
\rho_i = \frac{eB}{2\pi^2}\Bigg[{\sum_{s=\pm1}\sum_{\nu}\int_{0}^{\infty}  {{dk_{\parallel}}^i}\Big({{f^i}_{k,\nu,s}}-{\bar{f^i}_{k,\nu,s}}\Big)\Bigg]}, \label{charged_numberdensity}
\end{equation}
\begin{eqnarray}
 {\rho_s}^i = \frac{{eBm_i}^*}{2\pi^2}\Bigg[\sum_{s=\pm1}\sum_{\nu}\int_{0}^{\infty} \frac{\sqrt[]{{m_i}^{*2} + 2eB\nu} + s\Delta_i}{m_i^{*2} + 2eB\nu} \nonumber 
 \\\times  \frac {dk_{\parallel}^i}{\sqrt[]{{(k_{\parallel}^i)}^2+(\sqrt[]{m_i^{*2} + {2eB\nu}} + s\Delta_i)^2}}\Big({{f^i}_{k,\nu,s}}+{\bar{f^i}_{k,\nu,s}}\Big)   \Bigg]. 
 \label{charged_scalardensity}
\end{eqnarray}
Here $\nu$ is the Landau level, and spin index s = +1($-1$) corresponds to spin up (spin down) projections  for the baryons. The Landau levels of charged baryons are enumerated using the expression  $\nu = n+\frac{1}{2}-\frac{q_B}{|q_B|}\frac{s}{2}$ where $q_B$ is the charge of the baryon($q_B=e$ for p, $\Sigma^{+}$, and $q_B=-e$ for $\Sigma^{-}$, $\Xi^{-}$). The lowest Landau level for a particular spin projection of the charged baryon is obtained by setting n=0 in this expression. In the above equations, $k_{\parallel}^{i}$ denotes the momenta of baryons along the direction of the magnetic field and $m_i^* = - (g_{\sigma i}\sigma+ g_{\zeta i}\zeta + g_{\delta i}\delta$) is the effective mass of the baryons. Here $g_{\sigma i}, g_{\zeta i}, g_{\delta i}$ represent the coupling strengths of baryons with the scalar mesons. The parameter $\Delta_{i}=-\frac{1}{2}\kappa_i\mu_N B$  refers to the anomalous magnetic moments of the baryons. At finite temperature, ${f^i}_{k,\nu,s}$,  $\bar{f^i}_{k,\nu,s}$ are the particle and antiparticle temperature distribution functions,respectively, for charged baryons and given as
\begin{equation}
f^i_{k,\nu, s}= \frac{1}{1+\exp\left[\frac{( E^i_{\nu, s} 
-\mu^{*}_{i})}{T} \right]},\hspace{0.5cm}
\bar{f}^i_{k,\nu, s} = \frac{1}{1+\exp\left[\frac{( E^i_{\nu, s} 
+\mu^{*}_{i} )}{T}\right]}.
\label{chargedfermi_distribution}
\end{equation}

Here T is the temperature and $\mu_i^* =\mu_i-(g_{\rho i}\rho+g_{\omega i}\omega+g_{\phi i}\phi)$ is the effective chemical potential of baryons. $ E_{\nu, S}^i$ is the effective energy of charged baryons which is given as
\begin{eqnarray}
E_{\nu, S}^i = \sqrt[]{(k_{\parallel}^i)^2 + \Big(\sqrt[]{m_i^{*2} + 2eB\nu} + s\Delta_i\Big)^2}.
\label{charged_energy}
\end{eqnarray}
The number density and scalar density of neutral baryons ( i= $n$,$\Lambda$, $\Sigma^{0}$, $\Xi^{0}$ ) in an external magnetic field are given as
\begin{eqnarray}
\rho_i = \frac{1}{2\pi^2}\Bigg[\sum_{s=\pm1}\int_{0}^{\infty}{k^i_{\bot}} {dk^i_{\bot}}\int_{0}^{\infty}  dk_{\parallel}^i\Big({{f^i}_{k,s}}-{\bar{f^i}_{k,s}}\Big)   \Bigg],
\label{neutral_numberdensity}
\end{eqnarray}

\begin{eqnarray}
\rho_s^i = \frac{1}{2\pi^2}\Bigg[\sum_{s=\pm1}\int_{0}^{\infty}{k^i_{\bot}} {dk^i_{\bot}}\Bigg(1+\frac{s\Delta_i}{\sqrt[]{m_i^{*2} +{k^i_{\bot}}^2}}\Bigg)\nonumber\\\times \int_{0}^{\infty}  \frac {dk_{\parallel}^i\times m_i^{*}}{\sqrt[]{({k_{\parallel}^i})^2+(\sqrt[]{m_i^{*2} + ({k_{\bot}^i})^2} + s\Delta_i)^2}}\Big({{f^i}_{k,s}}+{\bar{f^i}_{k,s}}\Big)   \Bigg].
\label{neutral_scalardensity}
\end{eqnarray}

Here $k_{\bot}^i$ is the momenta of neutral baryons perpendicular to the direction of the magnetic field. ${f^i}_{k,s}$ and $\bar{f^i}_{k,s}$ are the thermal distribution functions for neutral baryons given as
\begin{equation}
f^i_{k, s}= \frac{1}{1+\exp\left[\frac{( E^i_{s} 
-\mu^{*}_{i})}{T} \right]},\hspace{0.5cm}
\bar{f}^i_{k, s} = \frac{1}{1+\exp\left[\frac{( E^i_{s} 
+\mu^{*}_{i} )}{T}\right]},
\label{neutralfermi_distribution}
\end{equation}
where $E_{s}^i$ is the effective energy of the neutral baryons given as 
\begin{equation}
 E^{i}_{s}=\sqrt{\left(k^{i}_{\parallel}\right)^{2} +
\left(\sqrt{m^{* 2}_{i}+\left(k^{i}_{\bot}\right)^{2} }+s\Delta_i\right)^{2}}.
\label{neutralenergy}
\end{equation}
The hyperonic matter is considered in chemical and thermal equilibrium, which provides the necessary equations of constraints to solve the system. The equations of motion of scalar fields are then solved self consistently at different baryon density $\rho_B= \sum_i\rho_i$ for given values of magnetic fields, temperature, isospin asymmetry parameter $\eta= \frac{-\sum_i{I_{3i}\rho_i}}{\rho_B}$ and strangeness fraction $f_s= \frac{\sum_i|{S_i|\rho_i}}{\rho_B}$. Here $I_{3i}$ is the third component of isospin, and $S_i$ is the strangeness quantum number for the $i^{th}$ baryon. In the following section, we shall describe the interaction of open bottom mesons with magnetized hot strange hadronic matter and their medium modifications.

\section{MASSES OF OPEN BOTTOM MESONS IN HOT MAGNETIZED MATTER}

 The interaction Lagrangian density of $B$ and $\bar{B}$ mesons in the strange hadronic matter in the chiral effective model, is given as \cite{AM_Divakar_PRC_2015}
\allowdisplaybreaks
\begin{eqnarray}
{\cal L} ^{B}_{\rm int} =\left(\frac{-i}{8 f_{B}^{2}}\right)\left[3\left(\bar{p} \gamma^{\mu} p+\bar{n} \gamma^{\mu} n\right)\left(\left(\left(\partial_{\mu} B^{+}\right) B^{-}-B^{+}\left(\partial_{\mu} B^{-}\right)\right)\right.\right.\nonumber \left.+\left(\left(\partial_{\mu} B^{0}\right) \bar{B}^{0}-B^{0}\left(\partial_{\mu} \bar{B}^{0}\right)\right)\right) \nonumber\\
+\left(\bar{p} \gamma^{\mu} p-\bar{n} \gamma^{\mu} n\right)\left(\left(\left(\partial_{\mu} B^{+}\right) B^{-}-B^{+}\left(\partial_{\mu} B^{-}\right)\right)-\left(\left(\partial_{\mu} B^{0}\right) \bar{B}^{0}-B^{0}\left(\partial_{\mu} \bar{B}^{0}\right)\right)\right)
\nonumber\hspace{1.8cm}\\+2\left(\bar{\Lambda}^{0} \gamma^{\mu} \Lambda^{0}+\bar{\Sigma}^{0} \gamma^{\mu} \Sigma^{0}\right)\left(\left(\left(\partial_{\mu} B^{+}\right) B^{-}-B^{+}\left(\partial_{\mu} B^{-}\right)\right)+\left(\left(\partial_{\mu} B^{0}\right) \bar{B}^{0}-B^{0}\left(\partial_{\mu} \bar{B}^{0}\right)\right)\right) \nonumber\hspace{0.7cm}\\
+2\left(\bar{\Sigma}^{+} \gamma^{\mu} \Sigma^{+}+\bar{\Sigma}^{-} \gamma^{\mu} \Sigma^{-}\right)\left(\left(\left(\partial_{\mu} B^{+}\right) B^{-}-B^{+}\left(\partial_{\mu} B^{-}\right)\right)+\left(\left(\partial_{\mu} B^{0}\right) \bar{B}^{0}-B^{0}\left(\partial_{\mu} \bar{B}^{0}\right)\right)\right) \nonumber\hspace{0.35cm} \\
+2\left(\bar{\Sigma}^{+} \gamma^{\mu} \Sigma^{+}-\bar{\Sigma}^{-} \gamma^{\mu} \Sigma^{-}\right)\left(\left(\left(\partial_{\mu} B^{+}\right) B^{-}-B^{+}\left(\partial_{\mu} B^{-}\right)\right)-\left(\left(\partial_{\mu} B^{0}\right) \bar{B}^{0}-B^{0}\left(\partial_{\mu} \bar{B}^{0}\right)\right)\right)\nonumber \hspace{0.35cm}\\
+\left(\bar{\Xi}^{0} \gamma^{\mu} \Xi^{0}+\bar{\Xi}^{-} \gamma^{\mu} \Xi^{-}\right)\left(\left(\left(\partial_{\mu} B^{+}\right) B^{-}-B^{+}\left(\partial_{\mu} B^{-}\right)\right)+\left(\left(\partial_{\mu} B^{0}\right) \bar{B}^{0}-B^{0}\left(\partial_{\mu} \bar{B}^{0}\right)\right)\right)\nonumber\hspace{0.72cm} \\
\left.+\left(\bar{\Xi}^{0} \gamma^{\mu} \Xi^{0}-\bar{\Xi}^{-} \gamma^{\mu} \Xi^{-}\right)\left(\left(\left(\partial_{\mu} B^{+}\right) B^{-}-B^{+}\left(\partial_{\mu} B^{-}\right)\right)-\left(\left(\partial_{\mu} B^{0}\right) \bar{B}^{0}-B^{0}\left(\partial_{\mu} \bar{B}^{0}\right)\right)\right)\right] \nonumber\hspace{0.57cm}\\
+\frac{m_{B}^{2}}{2 f_{B}}\left[\left(\sigma^{\prime}+\sqrt{2} \zeta_{b}^{\prime}\right)\left(B^{+} B^{-}+B^{0} \bar{B}^{0}\right)+\delta\left(B^{+} B^{-}-B^{0} \bar{B}^{0}\right)\right]\nonumber\hspace{4.15cm} \\
+\left(\frac{-1}{f_{B}}\right)\left[\left(\sigma^{\prime}+\sqrt{2} \zeta_{b}^{\prime}\right)\left(\left(\partial^{\mu} B^{+}\right)\left(\partial_{\mu} B^{-}\right)+\left(\partial^{\mu} B^{0}\right)\left(\partial_{\mu} \bar{B}^{0}\right)\right)\right.
\nonumber\hspace{4.25cm}\\\left.+ \delta\left(\left(\partial^{\mu} B^{+}\right)\left(\partial_{\mu} B^{-}\right)-\left(\partial^{\mu} B^{0}\right)\left(\partial_{\mu} \bar{B}^{0}\right)\right)\right]\nonumber\hspace{5.6cm}\\
+\frac{d_{1}}{2 f_{B}^{2}}\left[\left(\bar{p} p+\bar{n} n+\bar{\Lambda}^{0} \Lambda^{0}+\bar{\Sigma}^{+} \Sigma^{+}+\bar{\Sigma}^{0} \Sigma^{0}+\bar{\Sigma}^{-} \Sigma^{-}+\bar{\Xi}^{0} \Xi^{0}+\bar{\Xi}^{-} \Xi^{-}\right)\right. \nonumber\hspace{2.8cm}\\\times
\left.\left(\left(\partial_{\mu} B^{+}\right)\left(\partial^{\mu} B^{-}\right)+\left(\partial_{\mu} B^{0}\right)\left(\partial^{\mu} \bar{B}^{0}\right)\right)\right] \nonumber\hspace{6.5cm}\\
+\frac{d_{2}}{4 f_{B}^{2}}\left[3(\bar{p} p+\bar{n} n)\left(\left(\partial_{\mu} B^{-}\right)\left(\partial^{\mu} B^{+}\right)+\left(\partial_{\mu} \bar{B}^{0}\right)\left(\partial^{\mu} B^{0}\right)\right)\right.\nonumber\hspace{5cm} \\
+(\bar{p} p-\bar{n} n)\left(\left(\partial_{\mu} B^{-}\right)\left(\partial^{\mu} B^{+}\right)-\left(\partial_{\mu} \bar{B}^{0}\right)\left(\partial^{\mu} B^{0}\right)\right) \nonumber\hspace{4.88cm}\\
+2\left(\bar{\Lambda}^{0} \Lambda^{0}+\bar{\Sigma}^{0} \Sigma^{0}\right)\left(\left(\partial_{\mu} B^{-}\right)\left(\partial^{\mu} B^{+}\right)+\left(\partial_{\mu} \bar{B}^{0}\right)\left(\partial^{\mu} B^{0}\right)\right)\nonumber \hspace{3.65cm}\\
+2\left(\bar{\Sigma}^{+} \Sigma^{+}+\bar{\Sigma}^{-} \Sigma^{-}\right)\left(\left(\partial_{\mu} B^{-}\right)\left(\partial^{\mu} B^{+}\right)+\left(\partial_{\mu} \bar{B}^{0}\right)\left(\partial^{\mu} B^{0}\right)\right)\nonumber\hspace{3.3cm} \\
+2\left(\bar{\Sigma}^{+} \Sigma^{+}-\bar{\Sigma}^{-} \Sigma^{-}\right)\left(\left(\partial_{\mu} B^{-}\right)\left(\partial^{\mu} B^{+}\right)-\left(\partial_{\mu} \bar{B}^{0}\right)\left(\partial^{\mu} B^{0}\right)\right)\nonumber\hspace{3.3cm} \\
+\left(\bar{\Xi}^{0} \Xi^{0}+\bar{\Xi}^{-} \Xi^{-}\right)\left(\left(\partial_{\mu} B^{-}\right)\left(\partial^{\mu} B^{+}\right)+\left(\partial_{\mu} \bar{B}^{0}\right)\left(\partial^{\mu} B^{0}\right)\right)\nonumber\hspace{3.71cm} \\
\left.+\left(\bar{\Xi}^{0} \Xi^{0}-\bar{\Xi}^{-} \Xi^{-}\right)\left(\left(\partial_{\mu} B^{-}\right)\left(\partial^{\mu} B^{+}\right)-\left(\partial_{\mu} \bar{B}^{0}\right)\left(\partial^{\mu} B^{0}\right)\right)\right].\hspace{3.62cm}
\label{L_Bmesons}
\end{eqnarray}

The interaction Lagrangian density of Bottom-strange ($B_s$) mesons in the strange hadronic medium is given as \cite{Divakar_AM_IntJMod_2014}
\begin{eqnarray}
{\cal L} ^{B_s}_{\rm int}=-\frac{i}{4 f_{B_{S}}^{2}} \left[\left(2\left(\bar{\Xi}^{0} \gamma^{\mu} \Xi^{0}+\bar{\Xi}^{-} \gamma^{\mu}\Xi^{-}\right)+\bar{\Lambda}^{0} \gamma^{\mu} \Lambda^{0}+\bar{\Sigma}^{+} \gamma^{\mu} \Sigma^{+}\right.\right.\nonumber\hspace{4cm}\\
+ \left.\left.\bar{\Sigma}^{0} \gamma^{\mu} \Sigma^{0}+\bar{\Sigma}^{-} \gamma^{\mu} 
 \Sigma^{-}\right)\left(\bar{B}_{S}^{0}\left(\partial_{\mu} B_{S}^{0}\right)-\left(\partial_{\mu} \bar{B}_{S}^{0}\right) B_{S}^{0}\right)\right] \nonumber\hspace{3.1cm}\\
+ \frac{m_{B_{S}}^{2}}{\sqrt{2} f_{B_{S}}}\left[\left(\zeta^{\prime}+\zeta_{b}^{\prime}\right)\left(\bar{B}_{S}^{0} B_{S}^{0}\right)\right] -\frac{\sqrt{2}}{f_{B_{S}}}\left[\left(\zeta^{\prime}+\zeta_{b}^{\prime}\right)\left(\left(\partial_{\mu} \bar{B}_{S}^{0}\right)\left(\partial^{\mu} B_{S}^{0}\right)\right)\right] \nonumber\hspace{2.3cm}\\
+\frac{d_{1}}{2 f_{B_{S}}^{2}}\left[\left(\bar{p} p+\bar{n} n+\bar{\Lambda}^{0} \Lambda^{0}+\bar{\Sigma}^{+} \Sigma^{+}+\bar{\Sigma}^{0} \Sigma^{0}\right.\right.\nonumber\hspace{5.85cm} \\
\left.\left.+\bar{\Sigma}^{-} \Sigma^{-}+\bar{\Xi}^{0} \Xi^{0}+\bar{\Xi}^{-} \Xi^{-}\right)\left(\left(\partial_{\mu} \bar{B}_{S}^{0}\right)\left(\partial^{\mu} B_{S}^{0}\right)\right)\right] \nonumber\hspace{4.2cm}\\
+\frac{d_{2}}{2 f_{B_{S}}^{2}}\left[\left(2\left(\bar{\Xi}^{0} \Xi^{0}+\bar{\Xi}^{-} \Xi^{-}\right)+\bar{\Lambda}^{0} \Lambda^{0}+\bar{\Sigma}^{+} \Sigma^{+}\right.\right.\nonumber\hspace{5.55cm} \\
\left.\left.+\bar{\Sigma}^{0} \Sigma^{0}+\bar{\Sigma}^{-} \Sigma^{-}\right)\left(\left(\partial_{\mu} \bar{B}_{S}^{0}\right)\left(\partial^{\mu} B_{S}^{0}\right)\right)\right].\hspace{5.65cm}
\label{L_Bs}
\end{eqnarray}

The first term in eq.(\ref{L_Bmesons}) (with coefficient $-i/8{f_B}^2$) and the first term in eq.(\ref{L_Bs}) (with coefficient $-i/4{f_{B_s}}^2$) is the vectorial Weinberg-Tomozawa interaction term obtained from the kinetic energy term $\mathcal{L_\textrm{kin}}$ of eq.(\ref{genlag}) \cite{AM_Divakar_PRC_2015,Divakar_AM_IntJMod_2014}. The second term in eq.(\ref{L_Bmesons}) (with coefficient $m^2_B/2f_B$) and that in eq.(\ref{L_Bs}) (with coefficient $m^2_{B_S}/\sqrt{2}f_{B_s}$ ) is the scalar meson exchange term obtained from the explicit symmetry-breaking term $\mathcal{L_\textrm{SB}}$ in eq.(\ref{genlag}). The next three terms in eq.(\ref{L_Bmesons}) and  in eq.(\ref{L_Bs}) are known as the range terms. The first range term in eq.(\ref{L_Bmesons}) (with the coefficient ($- 1/ f_B$)) and that in eq.(\ref{L_Bs}) (with the coefficient ($-\sqrt{2}/ f_{B_s}$))) is obtained from the kinetic energy term of the pseudoscalar mesons. Here $f_B$ and $f_{B_s}$ refer to the decay constants of $B$ and $B_s$ mesons, respectively. The parameters $d_1$ and $d_2$ in the last two range terms of eq.(\ref{L_Bmesons}) and eq.(\ref{L_Bs})  are determined by a fit of the empirical values of the Kaon-Nucleon scattering lengths\cite{Brown,Bielich,Barnes_PRC49} for I = 0 and I = 1 channels\cite{AM_PRC78_2008, AM_EURPHYS_2009}. The interaction Lagrangian density gives rise to equations of motion for $B$, $\bar{B}$, and $B_s$ mesons and their Fourier transforms lead to the dispersion relations given as
\begin{eqnarray}
-\omega^2 + \overrightarrow{k}^2 + m_{j}^2 - \Pi_{j}(\omega,|\overrightarrow{k}|) = 0.
\label{disp_relation}
\end{eqnarray}

Here the index j denotes the various open bottom mesons $B$, $\bar{B}$, $B_s$, and $ m_{j}$ is the vacuum mass of the corresponding open bottom mesons. $\Pi_{j}(\omega,|\vec{k}|)$ denotes the self-energy of these mesons in the medium. For $B$ mesons, the self-energy is given as
\begin{eqnarray}
\Pi_{{B}} (\omega, |\vec k|) =\frac{-1}{4 f_{B}^{2}}\left[3\left(\rho_{p}+\rho_{n}\right) \pm\left(\rho_{p}-\rho_{n}\right)+2 \rho_{\Lambda}+2 \rho_{\Sigma^{0}}+2\left(\rho_{\Sigma^{+}}+\rho_{\Sigma^{-}}\right)\right.\nonumber\hspace{1.7cm} \\
\left.\pm 2\left(\rho_{\Sigma^{+}}-\rho_{\Sigma^{-}}\right)+\left(\rho_{\Xi^{0}}+\rho_{\Xi^{-}}\right) \pm\left(\rho_{\Xi^{0}}-\rho_{\Xi^{-}}\right)\right] \omega \nonumber\hspace{2.8cm}\\
+\frac{m_{B}^{2}}{2 f_{B}}\left(\sigma^{\prime}+\sqrt{2} \zeta_{b}^{\prime} \pm \delta^{\prime}\right) \nonumber\hspace{7.5cm}\\
+\left[\frac{d_{1}}{2 f_{B}^{2}}\left(\rho_{p}^{s}+\rho_{n}^{s}+\rho_{\Lambda}^{s}+\rho_{\Sigma^{+}}^{s}+\rho_{\Sigma^{0}}^{s}+\rho_{\Sigma^{-}}^{s}+\rho_{\Xi^{0}}^{s}+\rho_{\Xi^{-}}^{s}\right)\right. \nonumber\hspace{2.1cm}\\
+\frac{d_{2}}{4 f_{B}^{2}}\left(3\left(\rho_{p}^{s}+\rho_{n}^{s}\right) \pm\left(\rho_{p}^{s}-\rho_{n}^{s}\right)+2 \rho_{\Lambda}^{s}+2\left(\rho_{\Sigma^{+}}^{s}+\rho_{\Sigma^{-}}^{s}\right)\right. \nonumber\hspace{1.8cm}\\
\left.\pm 2\left(\rho_{\Sigma^{+}}^{s}-\rho_{\Sigma^{-}}^{s}\right)+2 \rho_{\Sigma^{0}}^{s}+\left(\rho_{\Xi^{0}}^{s}+\rho_{\Xi^{-}}^{s}\right) \pm\left(\rho_{\Xi^{0}}^{s}-\rho_{\Xi^{-}}^{s}\right)\right)\nonumber\hspace{0.7cm} \\
\left.-\frac{1}{f_{B}}\left(\sigma^{\prime}+\sqrt{2} \zeta_{b}^{\prime} \pm \delta^{\prime}\right)\right]\left(\omega^{2}-|\vec{k}|^{2}\right),\hspace{5cm}
\label{selfenergy_BplusB0}
\end{eqnarray}

where the $+$ and $-$ signs refer to the $B^+$  and $B^0$ mesons, respectively. For the  $\bar{B}$ meson doublet, the expression for self-energy is given as, 
\begin{eqnarray}
\Pi_{{\bar{B}}} (\omega, |\vec k|) =\frac{1}{4 f_{B}^{2}}\left[3\left(\rho_{p}+\rho_{n}\right) \pm\left(\rho_{p}-\rho_{n}\right)+2 \rho_{\Lambda}+2 \rho_{\Sigma^{0}}+2\left(\rho_{\Sigma^{+}}+\rho_{\Sigma^{-}}\right)\right.\nonumber\hspace{1.7cm} \\
\left.\pm 2\left(\rho_{\Sigma^{+}}-\rho_{\Sigma^{-}}\right)+\left(\rho_{\Xi^{0}}+\rho_{\Xi^{-}}\right) \pm\left(\rho_{\Xi^{0}}-\rho_{\Xi^{-}}\right)\right] \omega \nonumber\hspace{2.8cm}\\
+\frac{m_{B}^{2}}{2 f_{B}}\left(\sigma^{\prime}+\sqrt{2} \zeta_{b}^{\prime} \pm \delta^{\prime}\right) \nonumber\hspace{7.5cm}\\
+\left[\frac{d_{1}}{2 f_{B}^{2}}\left(\rho_{p}^{s}+\rho_{n}^{s}+\rho_{\Lambda}^{s}+\rho_{\Sigma^{+}}^{s}+\rho_{\Sigma^{0}}^{s}+\rho_{\Sigma^{-}}^{s}+\rho_{\Xi^{0}}^{s}+\rho_{\Xi^{-}}^{s}\right)\right. \nonumber\hspace{2.1cm}\\
+\frac{d_{2}}{4 f_{B}^{2}}\left(3\left(\rho_{p}^{s}+\rho_{n}^{s}\right) \pm\left(\rho_{p}^{s}-\rho_{n}^{s}\right)+2 \rho_{\Lambda}^{s}+2\left(\rho_{\Sigma^{+}}^{s}+\rho_{\Sigma^{-}}^{s}\right)\right. \nonumber\hspace{1.8cm}\\
\left.\pm 2\left(\rho_{\Sigma^{+}}^{s}-\rho_{\Sigma^{-}}^{s}\right)+2 \rho_{\Sigma^{0}}^{s}+\left(\rho_{\Xi^{0}}^{s}+\rho_{\Xi^{-}}^{s}\right) \pm\left(\rho_{\Xi^{0}}^{s}-\rho_{\Xi^{-}}^{s}\right)\right)\nonumber\hspace{0.7cm} \\
\left.-\frac{1}{f_{B}}\left(\sigma^{\prime}+\sqrt{2} \zeta_{b}^{\prime} \pm \delta^{\prime}\right)\right]\left(\omega^{2}-|\vec{k}|^{2}\right),\hspace{5cm}
\label{selfenergy_BminusBbar0}
\end{eqnarray}
where the $+$ and $-$ signs refer to $B^-$ and $\bar{B^0}$ mesons, respectively. The expression for self-energy for Bottom-strange  mesons reads
\begin{eqnarray}
\Pi_{{B_s}}(\omega,|\vec{k}|)=&\left[\left(\frac{d_{1}}{2 f_{B_{S}}^{2}}\left(\rho_{p}^{s}+\rho_{n}^{s}+\rho_{\Lambda}^{s}+\rho_{\Sigma^{+}}^{s}+\rho_{\Sigma^{0}}^{s}+\rho_{\Sigma^{-}}^{s}+\rho_{\Xi^{0}}^{s}+\rho_{\Xi^{-}}^{s}\right)\right)\right.\nonumber\hspace{1.3cm}\\
&+\left(\frac{d_{2}}{2 f_{B_{S}}^{2}}\left(2\left(\rho_{\Xi^{0}}^{s}+\rho_{\Xi^{-}}^{s}\right)+\rho_{\Lambda}^{s}+\rho_{\Sigma^{+}}^{s}+\rho_{\Sigma^{0}}^{s}+\rho_{\Sigma^{-}}^{s}\right)\right) \nonumber\hspace{1.5cm}\\
&\left.-\left(\frac{\sqrt{2}}{f_{B_{S}}}\left(\zeta^{\prime}+\zeta_{b}^{\prime}\right)\right)\right]\left(\omega^{2}-\vec{k}^{2}\right) \nonumber\hspace{5.3cm}\\
& \pm\left[\frac{1}{2 f_{B_{S}}^{2}}\left(2\left(\rho_{\Xi^{0}}+\rho_{\Xi^{-}}\right)+\rho_{\Lambda}+\rho_{\Sigma^{+}}+\rho_{\Sigma^{0}}+\rho_{\Sigma^{-}}\right)\right] \omega \nonumber\hspace{1.9cm}\\
&+\left[\frac{m_{B_{S}}^{2}}{\sqrt{2} f_{B_{S}}}\left(\zeta^{\prime}+\zeta_{b}^{\prime}\right)\right],\hspace{7cm}
\label{selfenergy_Bs)}
\end{eqnarray}

where the $+$ and $-$ signs refer to $\bar{B_s}^0$ and ${B_s}^0$  mesons, respectively. In eq.(\ref{selfenergy_BplusB0}), eq.(\ref{selfenergy_BminusBbar0}) and  eq.(\ref{selfenergy_Bs)}), $\sigma^\prime$ = ($\sigma - \sigma_0$), $\zeta^\prime_b$ = ($\zeta_b - \zeta_{b0}$), $\delta^\prime$ = ($\delta - \delta_0$), and $\zeta^\prime$ = ($\zeta - \zeta_0$) are the corresponding fluctuations of the scalar fields in the hot magnetized medium from their vacuum values. The fluctuation  $\zeta^\prime_b$ has been neglected in the current investigation since  we treat the bottom degrees of freedom to be frozen in the medium \cite{AM_Divakar_PRC_2015, Dhale}. The charged open bottom mesons ( j= $B^+$, $B^-$) have an additional positive mass modification in magnetic fields, which retaining only the lowest Landau level, is given as
\begin{eqnarray}
m^{eff}_{j} = \sqrt[]{m_{j}^{*2} + |eB|}.
\label{charged_effective_mass}
\end{eqnarray}
Here $m_{j}^{*}$ are solutions of the dispersion relations given by eq.(\ref{disp_relation}) for $\omega$ at $|\overrightarrow{k}|$ = 0. For the neutral open bottom mesons ( j= $B^0$, $\bar{B^0}$,${B_s}^0$, $\bar{{B_s}^0}$), there is no contribution from the landau quantization effects. Hence their effective mass in the medium are solutions of dispersion relation and given as, 
\begin{eqnarray}
m^{eff}_{j} = {m_{j}}^{*}. \label{neutral_effective_mass}
\end{eqnarray}

\section{MEDIUM MASS SHIFTS OF UPSILON STATES }
The masses of the upsilon states modify in the magnetized hadronic medium due to the medium modification of the gluon condensates. Due to scale symmetry breaking, the trace of the energy-momentum tensor in QCD is non-zero and will have contributions from gluon condensates and finite quark mass contributions. In the chiral effective model, the scale symmetry breaking is introduced via the scale breaking Lagrangian  given as
\begin{equation}
{\cal L}_{scalebreak}= -\frac{1}{4} \chi^{4} 
{\rm ln} \frac{\chi^{4}}{\chi_{0}^{4}} + \frac{d}{3} \chi^{4} {\rm ln} \Bigg( \frac{\left( \sigma^{2} - \delta^{2}\right)\zeta }
{\sigma_{0}^{2} \zeta_{0}} \Big( \frac{\chi}{\chi_{0}}\Big) ^{3}\Bigg). 
\label{Lscalebreak}
\end{equation}

The relation between the scalar gluon condensate and the dilaton field $\chi$ is obtained by equating the trace of energy-momentum tensor from QCD and from the scale breaking Lagrangian ${\cal L}_{scalebreak}$ \cite{Cohen}. Using the one-loop QCD beta function, with $N_c=3$ and $N_f$=3, the resulting expression is \cite{AM_Divakar_PRC_2014, Magstrange}
\begin{equation}
\left\langle  \frac{\alpha_{s}}{\pi} G_{\mu\nu}^{a} G^{ \mu\nu a} 
\right\rangle =  \frac{8}{9}\left[(1-d) \chi^{4}+\sum_{i} m_{i} \bar{q}_{i} q_{i}\right].
\label{chiglu}
\end{equation}
Here the parameter $d$ originates from the second logarithmic term in eq.(\ref{Lscalebreak}). The quark mass term term, $\sum_{i} m_i \bar q_i q_i$, which is related to the explicit chiral symmetry breaking term ${\cal L}_{SB}$ in eq.(\ref{genlag}) is given as \cite{Magstrange}
\begin{eqnarray}
\sum_i m_i \bar {q_i} q_i &=&  
\Big[ m_{\pi}^{2} 
f_{\pi} \sigma 
+ \left( \sqrt{2} m_{k}^{2}f_{k} - \frac{1}{\sqrt{2}} 
m_{\pi}^{2} f_{\pi} \right) \zeta  \Big].
\label{quarkmassterm} 
\end{eqnarray}
In the chiral effective model, the leading order mass shift formula of the upsilon states is then given as
\cite{Amal2_upsilon,AM_Divakar_PRC_2014}
\begin{equation}
\Delta m_{\Upsilon}=  \frac{1}{18}  \int dk^{2} 
\langle \vert \frac{\partial \psi (\vec k)}{\partial {\vec k}} 
\vert^{2} \rangle
\frac{k}{k^{2} / m_{b} + \epsilon} \times  \bigg ( 
\left\langle \frac{\alpha_{s}}{\pi} 
G_{\mu\nu}^{a} G^{\mu\nu a}\right\rangle -
\left\langle \frac{\alpha_{s}}{\pi} 
G_{\mu\nu}^{a} G^{\mu\nu a}\right\rangle _{0}
\bigg ),  
\label{massupsilon}
\end{equation}
where 
\begin{equation}
\langle \vert \frac{\partial \psi (\vec k)}{\partial {\vec k}} 
\vert^{2} \rangle
=\frac {1}{4\pi}\int 
\vert \frac{\partial \psi (\vec k)}{\partial {\vec k}} \vert^{2}
d\Omega.
\label{wavefunctionderivative}
\end{equation}
In eq.(\ref{massupsilon}) $m_b$ is the mass of the bottom quark,
$m_\Upsilon$ is the vacuum mass of the corresponding upsilon state and $\epsilon = 2 m_{b} - m_{\Upsilon}$ is the binding energy. Here, $ \langle \frac{\alpha_{s}}{\pi} 
G_{\mu\nu}^{a} G^{\mu\nu a}\rangle$ and $ \langle \frac{\alpha_{s}}{\pi} 
G_{\mu\nu}^{a} G^{\mu\nu a}\rangle_{0}$ are the expectation values of the scalar gluon condensates in the hot magnetized medium and in the vacuum, respectively.  The wave function of the upsilon state in the momentum space, denoted as
$\psi (\vec k)$ is  normalized as $\int\frac{d^{3}k}{(2\pi)^{3}} 
\vert \psi(\vec k) \vert^{2} = 1 $. 
The momentum space wave functions of the upsilon states are obtained through Fourier transformations of the coordinate space wave functions. They are taken to be harmonic oscillator wave functions \cite{Friman} and given as 
\begin{equation}
\psi_{Nl} (\vec r) = N_{N l}
\times 
(\beta^{2} r^{2})^{\frac{1}2{} l} \exp\Big({-\frac{1}{2} \beta^{2} r^{2}}
\Big) 
L_{N - 1}^{l + \frac{1}{2}} \left( \beta^{2} r^{2}\right)
Y_{lm} (\theta, \phi) 
\equiv R_{Nl} (r) Y_{lm}(\theta,\phi),
\label{wavefunction} 
\end{equation} 
where $\beta = \sqrt {M \omega / h}$ characterizes the strength of the 
harmonic potential with $M = m_{b}/2$ and
$L_p^ q \left( \beta^{2} r^{2}\right)$
is the associated Laguerre Polynomial.
The wave functions in the coordinate space are
normalized as $
\int d^3 r |\psi _{Nlm} (\vec r) |^2 =1$ with 
$\int _0 ^\infty |R_{Nl} (r)|^2 r^2 dr=1$, which
determines the normalization constants $N_{Nl}$. The spherical harmonics $Y_{lm} (\theta,\phi)$ satisfies the orthonormality condition 
$\int Y_{lm}(\theta,\phi)Y_{l'm'}(\theta,\phi)d\Omega 
=\delta_{ll'}\delta_{m m'}$.
The coordinate space wave functions for 1S, 2S, 3S, 4S, and 1D states are explicitly given as
\begin{eqnarray}
\psi_{1S} (r, \theta, \phi ) = \frac{N_{1S}}{ \sqrt{4 \pi }}  e^{-\frac{1}{2} \beta ^2r^2},
\label{1Swavefunction}
\end{eqnarray}
\begin{eqnarray}
\psi_{2S} (r, \theta, \phi ) = \frac{N_{2S}}{ \sqrt{4 \pi }}  e^{-\frac{1}{2} \beta ^2r^2}( \frac{3}{2}- \beta^2r^2),
\label{2Swavefunction}
\end{eqnarray}
\begin{eqnarray}
\psi_{3S} (r, \theta, \phi ) = \frac{N_{3S}}{ \sqrt{4 \pi }}  e^{-\frac{1}{2} \beta ^2r^2}( \frac{15}{8}- \frac{5}{2}\beta^2r^2+ \frac{1}{2} \beta^4r^4),
\label{3Swavefunction}\end{eqnarray}
 \begin{eqnarray}
  \psi_{4S} (r, \theta, \phi ) = \frac{N_{4S}}{ \sqrt{4 \pi }}  e^{-\frac{1}{2} \beta ^2r^2}( \frac{35}{16}- \frac{35}{8}\beta^2r^2+ \frac{7}{4} \beta^4r^4-\frac{1}{6}\beta^6r^6),
  \label{4Swavefunction}
 \end{eqnarray}
 \begin{eqnarray}
\psi_{1D} (r, \theta, \phi ) = {N_{1D}} {\beta^2r^2}    e^{-\frac{1}{2} \beta ^2r^2}\sqrt{\frac{5}{16 \pi}} (3 \cos^2 \theta -1).
\label{1Dwavefunction}
\end{eqnarray}

The strengths of the harmonic oscillator potential, $\beta$, for upsilons in 1S, 2S, 3S, and 4S states are calculated from their observed leptonic decay widths,
$\Gamma (\Upsilon(NS) \rightarrow e^+e^-$) in accord with the expression 
\cite{AM_Divakar_PRC_2014,Amal2_upsilon, Repko}
\begin{equation}
\Gamma (\Upsilon(NS) \rightarrow e^+e^-)
=
\frac{4 \alpha^2}{9 m_{\Upsilon (NS)}^2} 
|R_{NS}(r=0)|^2.
\label{leptonicdecaywidth}
\end{equation}
Here $\alpha=1/137$ is the fine structure constant, $m_{\Upsilon (NS)}$ is the vacuum mass of the corresponding upsilon state $\Upsilon(NS)$, and $R_{NS}(r=0)$ is the radial part of the wave function at the origin. Since the parameter $\beta$ is contained in the radial part of the wave function, their values for each state could be determined from the above expression of leptonic decay widths. For $\Upsilon(1D)$, the value of $\beta$ is obtained from the interpolation of the mass versus $\beta$ graph of the upsilon 1S, 2S, 3S, and 4S states. In the eq.(\ref{massupsilon}), the difference in the value of scalar gluon condensate in the medium and in the vacuum is given by,
\begin{eqnarray}
\bigg ( 
\left\langle \frac{\alpha_{s}}{\pi} 
G_{\mu\nu}^{a} G^{\mu\nu a}\right\rangle -
\left\langle \frac{\alpha_{s}}{\pi} 
G_{\mu\nu}^{a} G^{\mu\nu a}\right\rangle _{0}
\bigg )=\nonumber\hspace{6cm}\\\frac{8}{9}\left[(1-d)\left(\chi^{4}-\chi_{0}^{4}\right)+m_{\pi}^{2} f_{\pi} \sigma^{\prime}+\left(\sqrt{2} m_{K}^{2} f_{K}-\frac{1}{\sqrt{2}} m_{\pi}^{2} f_{\pi}\right) \zeta^{\prime}\right].
\label{massivegluoncondensate}
\end{eqnarray}

In the above equation, $\chi_0$ denotes the vacuum value of the dilaton field, and the terms proportional to $\sigma '$ and $\zeta '$ originate from the finite quark mass term $\sum_i m_i \bar {q_i} q_i$.

\section{RESULTS AND DISCUSSION}

From the chiral effective Lagrangian given in eq.(\ref{genlag}), after mean-field approximation,  we obtain the coupled equation of motion of scalar fields ($\sigma$, $\zeta$, $\delta$, and $\chi$) and for the vector fields ($\omega$, $\rho$, and $\phi$). The values of parameters of the chiral effective model and their fitting are given in Ref \cite{Magstrange}. The vacuum values of scalar fields denoted as $\sigma_0$, $\zeta_0$, and $\chi_0$  are $-93.3$ MeV, $-106.6$ MeV, and 409.8 MeV, respectively. The effects of temperature are incorporated through thermal distribution functions in the expression of number densities, scalar densities of baryons given by eq.(\ref{charged_numberdensity}), eq.(\ref{charged_scalardensity}), eq.(\ref{neutral_numberdensity}), and eq.(\ref{neutral_scalardensity}). The effect of magnetic fields appears through summation over the Landau energy levels and through AMM for charged baryons. The major contribution to the number density and scalar density of charged baryons comes from the lowest Landau level, especially at low temperatures. The contribution from higher Landau levels is also considered in the present investigation. However, their contributions are observed to be subdominant.

\begin{figure}[htbp]
\includegraphics[height=18.15cm, width=11cm, keepaspectratio=true]{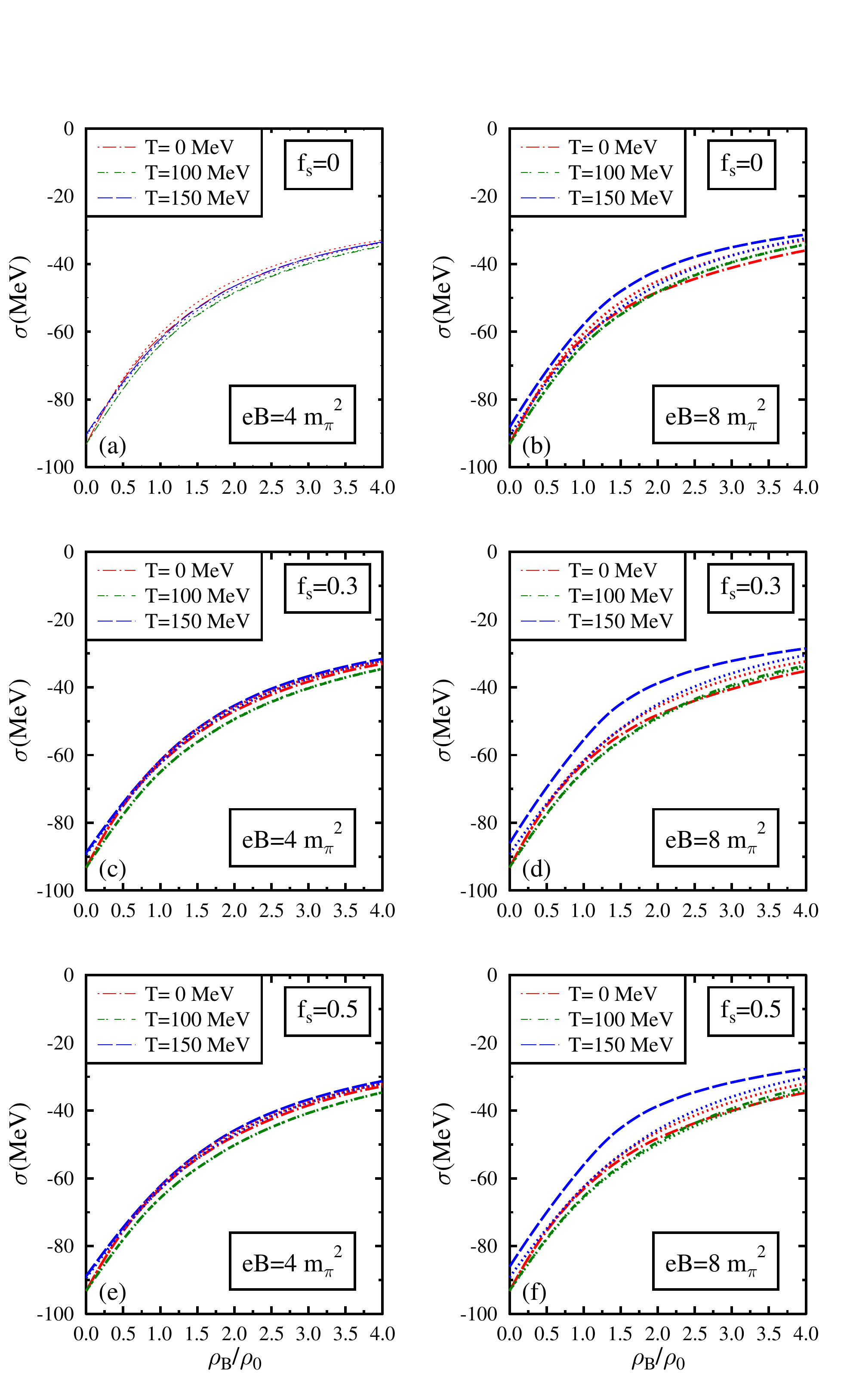}
\caption{The scalar field $\sigma$ in isospin asymmetric ($\eta$=0.5) hadronic matter is plotted as a function of the baryon density $\rho_{B} $/$\rho_{0}$  for different values of temperature T= 0, 100 and 150 MeV. These are plotted at magnetic fields $eB=4m_{\pi}^2$ and $eB=8m_{\pi}^2$, for a fixed value of strangeness fraction $f_s$ = 0, 0.3, 0.5. The effects of the anomalous magnetic moment of baryons are taken into account (dashed lines) and compared to the case when these effects are not considered (dotted lines).}
\label{sigma}
\end{figure}

\begin{figure}[htbp]
\includegraphics[height=18.15cm, width=11cm, keepaspectratio=true]{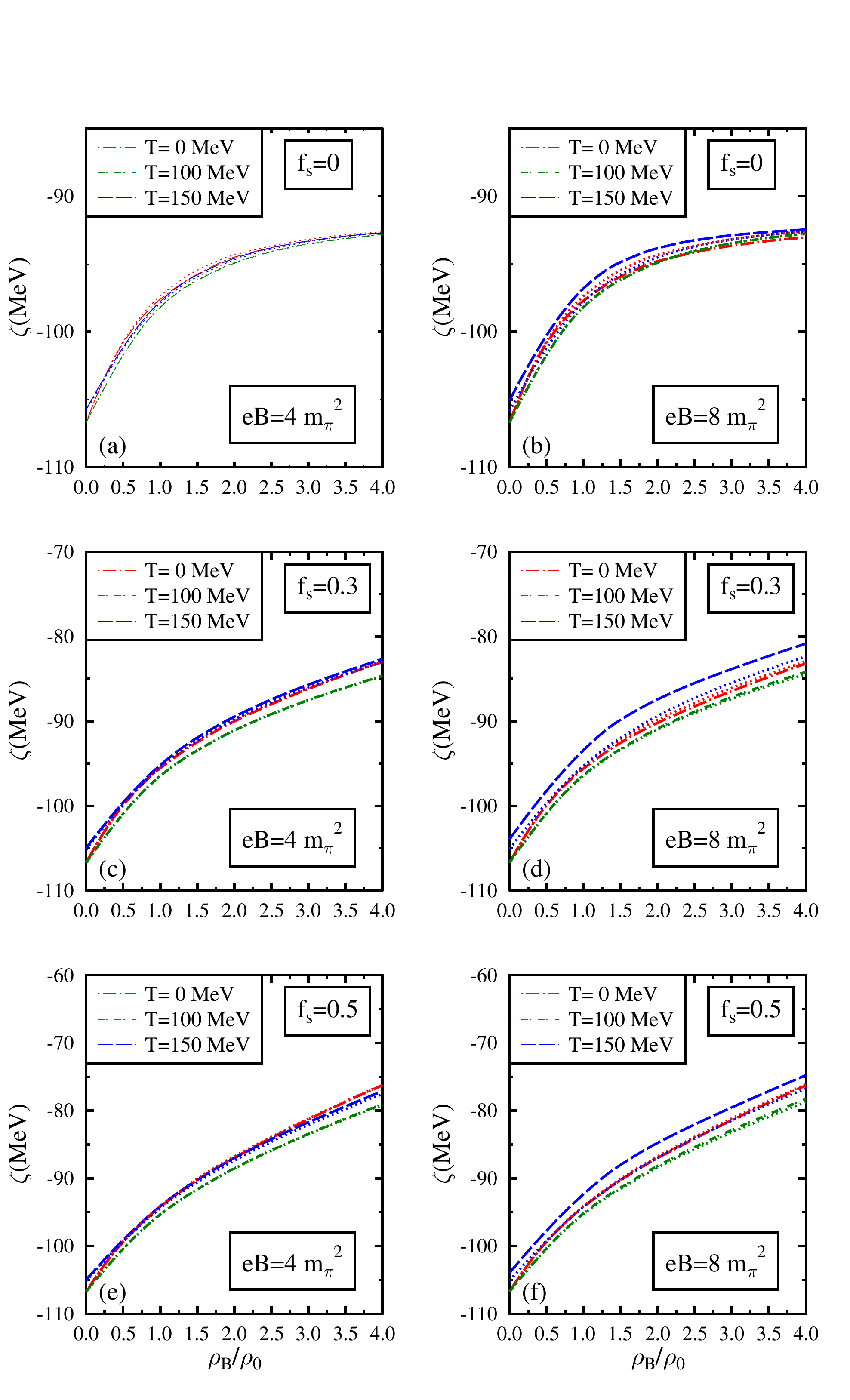}
\caption{The scalar field $\zeta$ in isospin asymmetric ($\eta$=0.5) hadronic matter is plotted as a function of the baryon density $\rho_{B} $/$\rho_{0}$  for different values of temperature T= 0, 100, and 150 MeV. These are plotted at magnetic fields $eB=4m_{\pi}^2$ and $eB=8m_{\pi}^2$, for a fixed value of strangeness fraction $f_s$ = 0, 0.3, 0.5. The effects of the anomalous magnetic moment of baryons are taken into account (dashed lines) and compared to the case when these effects are not considered (dotted lines).}
\label{zeta}
\end{figure}

\begin{figure}[htbp]
\includegraphics[height=18.15cm, width=11cm, keepaspectratio=true]{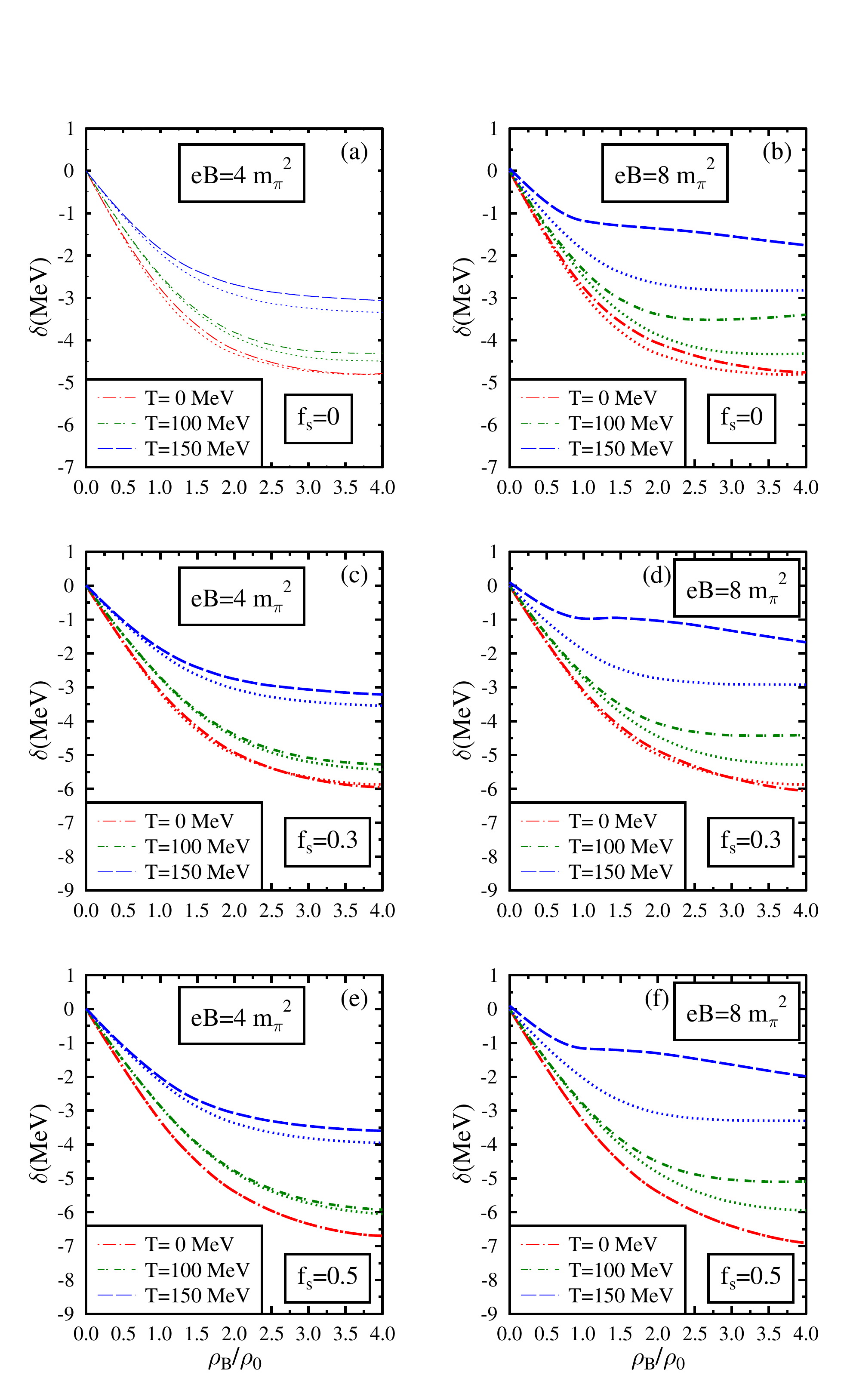}
\caption{The scalar field $\delta$ in isospin asymmetric ($\eta$=0.5) hadronic matter is plotted as a function of the baryon density $\rho_{B} $/$\rho_{0}$  for different values of temperature T= 0, 100, and 150 MeV. These are plotted at magnetic fields $eB=4m_{\pi}^2$ and $eB=8m_{\pi}^2$, for a fixed value of strangeness fraction $f_s$ = 0, 0.3, 0.5. The effects of the anomalous magnetic moment of baryons are taken into account (dashed lines) and compared to the case when these effects are not considered (dotted lines).}
\label{delta}
\end{figure}

\begin{figure}[htbp]
\includegraphics[height=18.15cm, width=11cm, keepaspectratio=true]{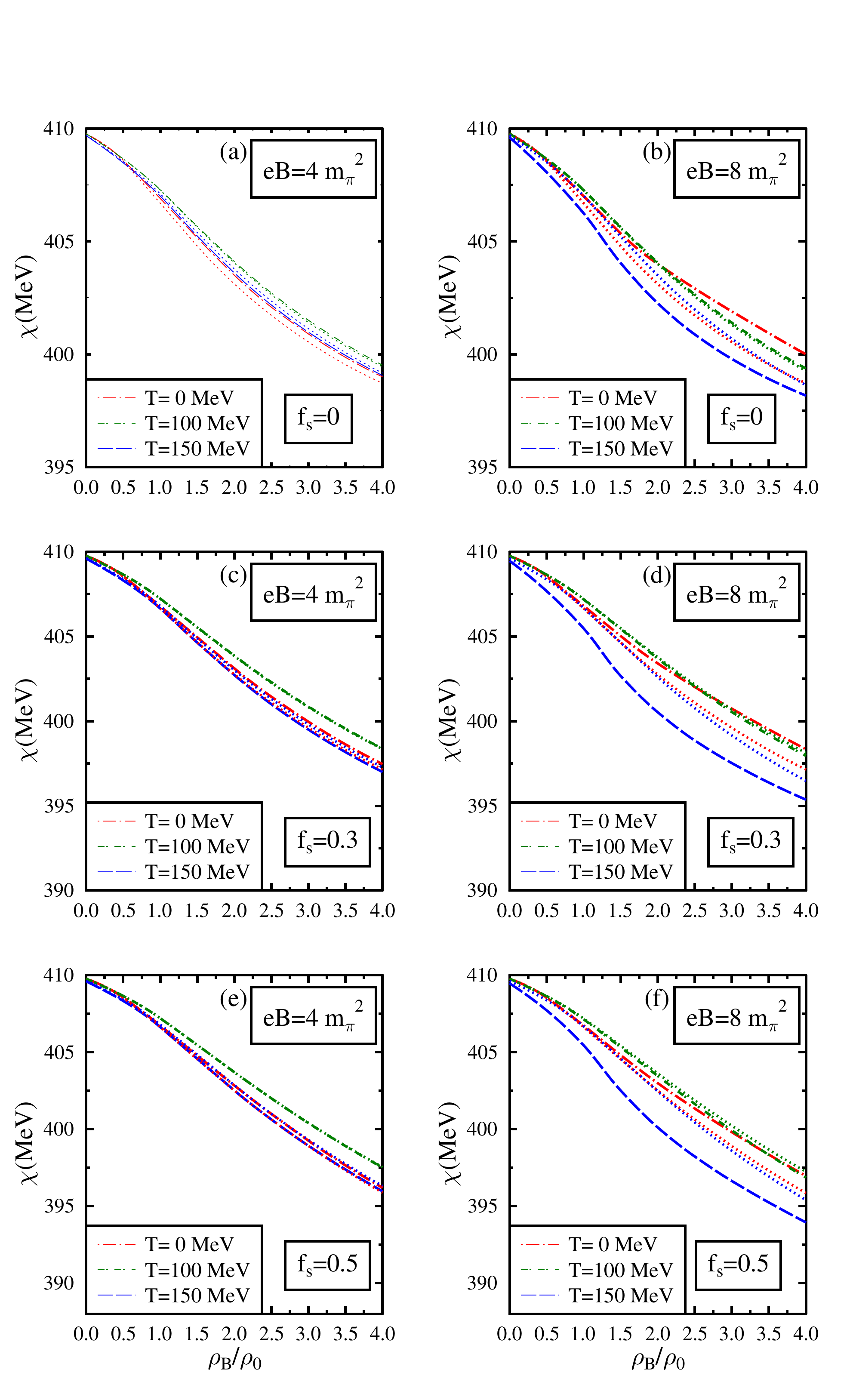}
\caption{The scalar field $\chi$ in isospin asymmetric ($\eta$=0.5) hadronic matter is plotted as a function of the baryon density $\rho_{B} $/$\rho_{0}$  for different values of temperature T= 0, 100, and 150 MeV. These are plotted at magnetic fields $eB=4m_{\pi}^2$ and $eB=8m_{\pi}^2$, for a fixed value of strangeness fraction $f_s$ = 0, 0.3, 0.5. The effects of the anomalous magnetic moment of baryons are taken into account (dashed lines) and compared to the case when these effects are not considered (dotted lines).}
\label{chi}
\end{figure}

The coupled equations of motion of  the scalar fields $\sigma$, $\zeta$, $\delta$ and $\chi$ are solved self consistently and are plotted in figs \ref{sigma}, \ref{zeta}, \ref{delta}, and \ref{chi} in the isospin asymmetric ($\eta=0.5$) hadronic matter as a function of baryon density $\rho_{B} $/$\rho_{0} $ where $\rho_{0} $ is nuclear matter saturation density. They are plotted for temperature T=0 MeV, 100 MeV, and 150 MeV  at magnetic fields  eB= $4{m_{\pi}}^2$ and $8{m_{\pi}}^2$. Each subplot is plotted for a particular value of strangeness fraction $f_{s}$=0 (corresponding to pure nuclear matter) and $f_{s}$= 0.3, 0.5 (corresponding to strange hadronic matter). The effects of the AMM of the baryons are also included in these plots and compared with the case where these effects are ignored (shown as dotted lines).

 For a fixed value of the magnetic field, strangeness fraction, and temperature, the magnitude of the scalar fields $\sigma$, $\zeta$,  and $\chi$ are observed to decrease with an increase in the baryon density. The value of $\zeta$ shows a saturation behavior in the magnetized nuclear matter at large densities. The modification of $\zeta$ is more significant in the magnetized strange hadronic matter as compared to the nuclear matter. The magnitude of $\delta$ initially rises with an increase in baryon density and then exhibits a saturation behavior at large densities in the nuclear matter as well as in the hyperonic matter. At large densities, as $f_s$ increases, the magnitude of $\sigma$, $\zeta$ and $\chi$ decreases whereas the magnitude of $\delta$ increases. The effects of AMM are prominent at larger magnetic fields and larger densities. The variation of scalar fields due to the magnetic field is small at T=0 MeV, especially at small densities. At T=0 MeV, for a given value of $f_s$  and density, the magnitude of the scalar fields $\sigma$, $\zeta$, and $\chi$ increases marginally with an increase in the magnetic field when AMM effects are taken into account. A detailed discussion of the qualitative and quantitative behavior of scalar fields in the hadronic matter as a function of baryon density, strangeness fraction, and magnetic fields at temperature T=0 has been given in Ref \cite{Magstrange}. The expression for number density and scalar density of baryons given in Ref \cite{Magstrange} are analytical expressions at T=0 MeV. They can be reproduced from the general expressions of number density and scalar density in terms of thermal distribution functions given by eq. (\ref{charged_numberdensity}), eq.(\ref{charged_scalardensity}), eq.(\ref{neutral_numberdensity}), and eq.(\ref{neutral_scalardensity}). In this paper, we focus on the effects of finite temperature and the magnetic field on the modification of scalar fields and subsequently on the in-medium masses of the open bottom mesons and upsilon states in the strange hadronic matter.

 At $\rho_B=0$, the magnitude of $\sigma$ and $\zeta$  decreases marginally as the temperature rises until T= 110-120 MeV and decreases sharply thereafter. The dilaton field $\chi$ exhibits similar behavior as its equation of motion is coupled with other scalar fields. However, the variation of $\chi$ with temperature is even smaller. The scalar densities of baryons can be non-zero even when their number density is zero. Hence at finite temperature, even when $\mu_i^*$=0, the value of the thermal distribution function increases with temperature, increasing the value of the scalar density of baryons. This effect of modification of thermal distribution function on scalar densities is weak till T= 110-120 MeV. Hence the modification of scalar fields with temperature is negligible in this regime. Above T= 120 MeV, the scalar densities and the scalar fields $\sigma$ and $\zeta$ undergo more significant modification as a function of temperature. The above behavior could be an indicator of the presence of baryon-antibaryon pairs in the thermal bath, which has been mentioned in the literature, without considering the presence of the magnetic fields \cite{AM_PRC69_2004, Furnstahl}. Consequently, the baryon masses will also get modified at high-temperature. The effect of temperature gets amplified with an increase in the magnetic field. 

 At finite baryon density, the magnitude of scalar fields $\sigma$, $\zeta$, and $\chi$ initially increase with an increase in temperature. At eB= $4{m_{\pi}}^2$, the magnitude of scalar fields at T= 100 MeV is larger than their magnitude at T =0 MeV. This behavior is due to the contributions from higher momenta at finite densities, which reduces the integrand in the equation for scalar density given in eq.(\ref{charged_scalardensity}) and eq.(\ref{neutral_scalardensity}). These contributions result in a drop in scalar density and increase the value of scalar fields with an increase in temperature. However, beyond a particular temperature, the magnitude of these scalar field decreases as a function of temperature due to the significant increase in the value of thermal distribution functions. Consequently, there is a net increase in the value of scalar density of baryons. For $f_s$=0 and eB= $4{m_{\pi}}^2$, the corresponding turnaround temperature is approximately T= 110 MeV at nuclear matter saturation density. This value of turnaround temperature decreases significantly with an increase in density and when the magnetic field becomes eB= $8{m_{\pi}}^2$. Here the magnitude of scalar fields initially rises with temperature, attaining their maximum values in the temperature regime T= 50-80 MeV, and thereafter their magnitude drops. As a result, at eB= $8{m_{\pi}}^2$, the magnitude of scalar fields at T= 100 MeV is smaller than the magnitude at T =0 MeV when  the density is above $\rho_B$=2$\rho_0$. It is observed that this drop in the magnitude of scalar fields at high temperatures is more significant at larger magnetic fields and strangeness fractions. 

At $\rho_B=0$, the effect of the magnetic field on the scalar fields is negligible below T=100 MeV. Above T=100 MeV, when the effects of AMM of baryons are incorporated, the magnitude of scalar fields $\sigma$, $\zeta$ and $\chi$ drops as the magnetic field is increased from eB= $4{m_{\pi}}^2$ to eB= $8{m_{\pi}}^2$. This behavior is due to the contributions of the non-zero scalar density of baryons at high temperatures, even when $\rho_B=0$, which gets modified by the magnetic field. The effect of magnetic fields on the scalar fields is more significant at high temperatures than at low temperatures. 

In asymmetric nuclear matter, when AMM effects are taken into account, the magnitude of the scalar fields $\sigma$, $\zeta$, and $\chi$ increases as the strength of the magnetic field increases when T$<$90 MeV. This behavior is because the scalar density of neutrons (${\rho_n}^s$) decreases with an increase in the magnetic field at moderate temperature. This positive modification of the magnitude of scalar fields by the magnetic field becomes small with a rise in temperature. Above T=90 MeV, the magnitude of scalar fields instead drops with an increase in the magnetic field. For asymmetric nuclear matter, the contribution of the scalar density of protons (${\rho_p}^s$) to the scalar fields becomes significant when T$>$ 90 MeV. Consequently, the variation of ${\rho_p}^s$ with a change in the magnetic field also becomes significant when T$>$ 90 MeV. In this temperature regime, the positive modification of ${\rho_p}^s$ with an increase in the magnetic field overshoots the corresponding negative modification of ${\rho_n}^s$. This net drop in the value of scalar fields becomes quite significant at T=150 MeV. Thus the effect of the magnetic field on the scalar fields, which mimics the condensates of QCD, is different at low temperatures and high temperatures in the magnetized strange hadronic matter. Such behavior for chiral condensates has been well documented in the literature at zero baryon density \cite{Balimag, Balimag2} and at finite baryon density \cite{Holographic, Rajeshkumar, Rajeshkumar2}. The presence of the magnetic field at high temperature may enhance the restoration of chiral symmetry and may result in the decrease of the critical temperature \cite{Balimag, Balimag2, Rajeshkumar, Rajeshkumar2}. The effects of AMM of baryons on these scalar fields become quite significant at large densities, larger magnetic fields, and higher temperatures.

The scalar-isovector field $\delta$ is zero at $\rho_B=0$ in the absence of the magnetic field. The value of $\delta$ becomes non-zero at $\rho_B=0$ only at large magnetic fields and high temperatures. For eB= $8{m_{\pi}}^2$, at T=150 MeV, when the effects of AMM are incorporated, the value of $\delta$ becomes marginally positive.  At T=150 MeV, the scalar densities for baryons are non-zero even though their number densities are zero. The scalar density of different baryons is modified differently in the presence of the magnetic field due to the difference in their charge and the values of their  AMM. The modifications of scalar densities are profound at high temperatures and large magnetic fields. Since $\delta$ is proportional to the difference in the scalar densities of the isospin partners, $\delta$ will acquire a nonzero value. Similar behavior of $\delta$ field at finite temperature and the magnetic field is also obtained in Ref \cite{Rajeshkumar}. At finite density, for $\eta=0.5$, the value of $\delta$ becomes negative, and its magnitude decreases with progressively increasing temperature. For $f_s$=0, the scalar density of protons increases, and the scalar density of neutrons decreases with an increase in temperature, resulting in a smaller value of ${\rho_n}^s$- ${\rho_p}^s$. Hence we observe a  decline in the magnitude of $\delta$ as a function of temperature. This effect is dominant in large magnetic fields when the effects of AMM are incorporated. The magnitude of $\delta$ considerably increases as a function of baryon density at low temperatures and then saturates at large baryon densities. At T=150 MeV, the saturation occurs at relatively small densities, as shown in Fig(\ref{delta}). In strange hadronic matter, $\delta$ undergoes larger modification. However, when the temperature is increased to T=150 MeV, the effect of $f_s$ on $\delta$ is small, and the effect of the magnetic field is significant. The magnitude of $\delta$ is larger when the effect of AMM on baryons is neglected.

 For $f_s=0.3$, when AMM effects are incorporated, the values of $\sigma$ in MeV at $\rho_B$=1$\rho_0$(4$\rho_0$)  are $-62.42$ $(-33.08)$, $-65.07$ $(-34.56)$, and $-61.63$ $(-31.59)$ for T=0, 100, and 150 MeV, respectively, at eB=$4m_\pi^2$. For eB=$8m_\pi^2$, under similar conditions, the values of $\sigma$ are  $-62.57$ $(-35.17)$, $-64.76$ $(-33.51)$, and $-55.52$ $(-28.53)$. For eB=$4m_\pi^2$, the values of $\zeta$ in MeV at $\rho_B$=1$\rho_0$(4$\rho_0$) are observed to be $-95.55$ $(-83.01)$, $-96.47$ $(-84.63)$,  $-95.19$ $(-82.66)$ for T=0, 100, and 150 MeV, respectively. For eB=$\ 8m_\pi^2$, they become $-95.58$ $(-83.18)$, $-96.35$($-84.17$), and $-93.40$ $(-80.84)$. At eB=$4m_\pi^2$, the values of $\delta$ in MeV under similar medium conditions are $-3.10$ $(-5.94)$, $-2.70$ $(-5.27)$, and $-1.85$ $(-3.21)$ . When the magnetic field is increased to  eB=$8m_\pi^2$, the values of $\delta$ in the same order are  $-3.10$ $(-6.05)$, $-2.63$ $(-4.40)$, and $-0.97$  $(-1.67)$.  At $\rho_B$=1$\rho_0$(4$\rho_0$), eB= $4m_\pi^2$, the dilaton field $\chi$ takes the values 406.78 (397.46), 407.23 (398.37), 406.65 (397.00) for T=0, 100, and 150 MeV,respectively. For eB= $8m_\pi^2$, these values in MeV become 406.80 (398.34), 407.18 (397.95), and 405.47 (395.36), respectively.

\begin{figure}[htbp]
\includegraphics[height=18.15cm, width=11cm, keepaspectratio=true]{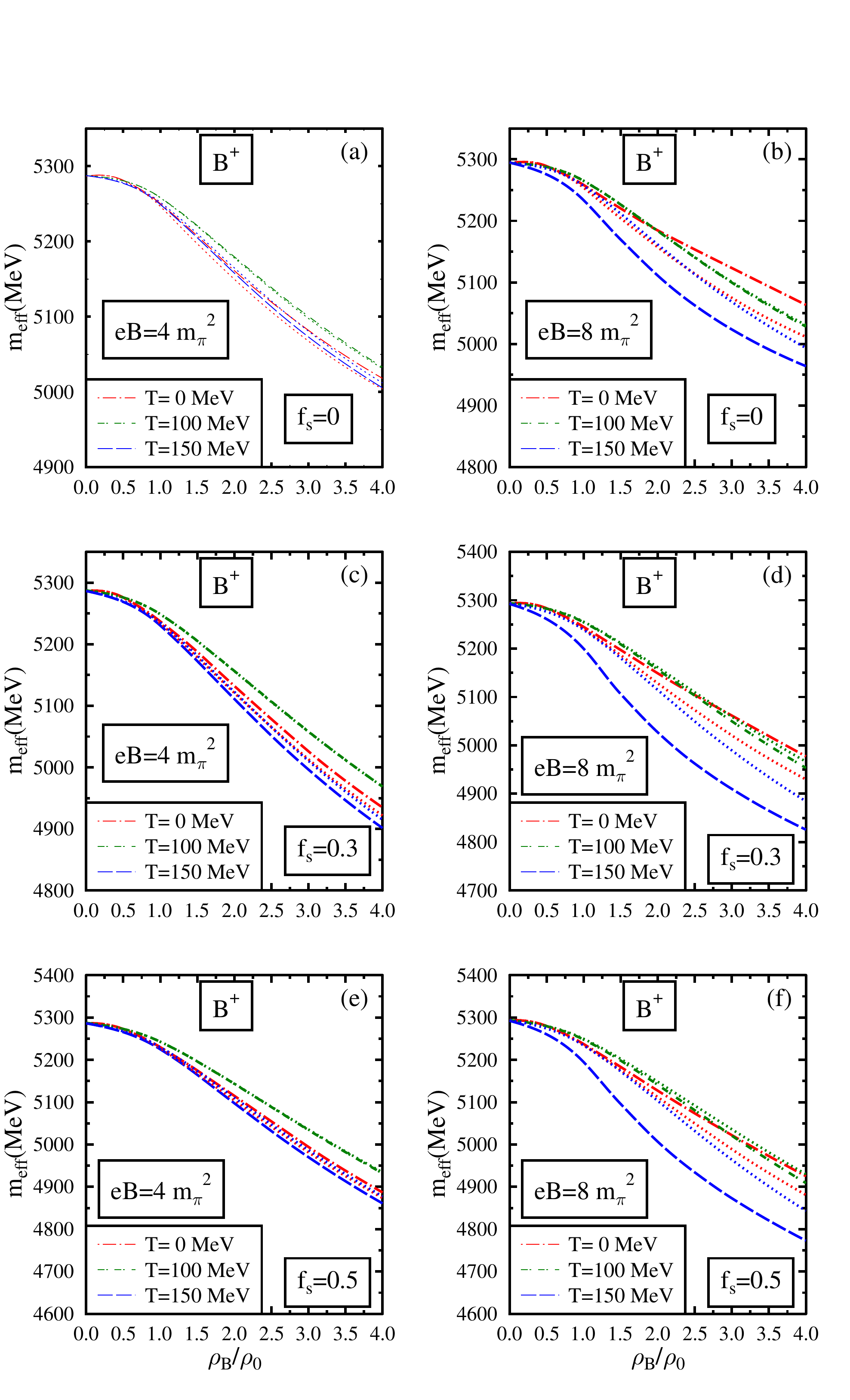}
\caption{The effective mass of $B^+$ meson in isospin asymmetric ($\eta$=0.5) hadronic matter is plotted as a function of the baryon density $\rho_{B} $/$\rho_{0}$  for different values of temperature T= 0, 100, and 150 MeV. These are plotted at magnetic fields $eB=4m_{\pi}^2$ and $eB=8m_{\pi}^2$, for a fixed value of strangeness fraction $f_s$ = 0, 0.3, 0.5. The effects of the anomalous magnetic moment of baryons are taken into account (dashed lines) and compared to the case when these effects are not considered (dotted lines).}
\label{mBplus}
\end{figure}

\begin{figure}[htbp]
\includegraphics[height=18.15cm, width=11cm, keepaspectratio=true] {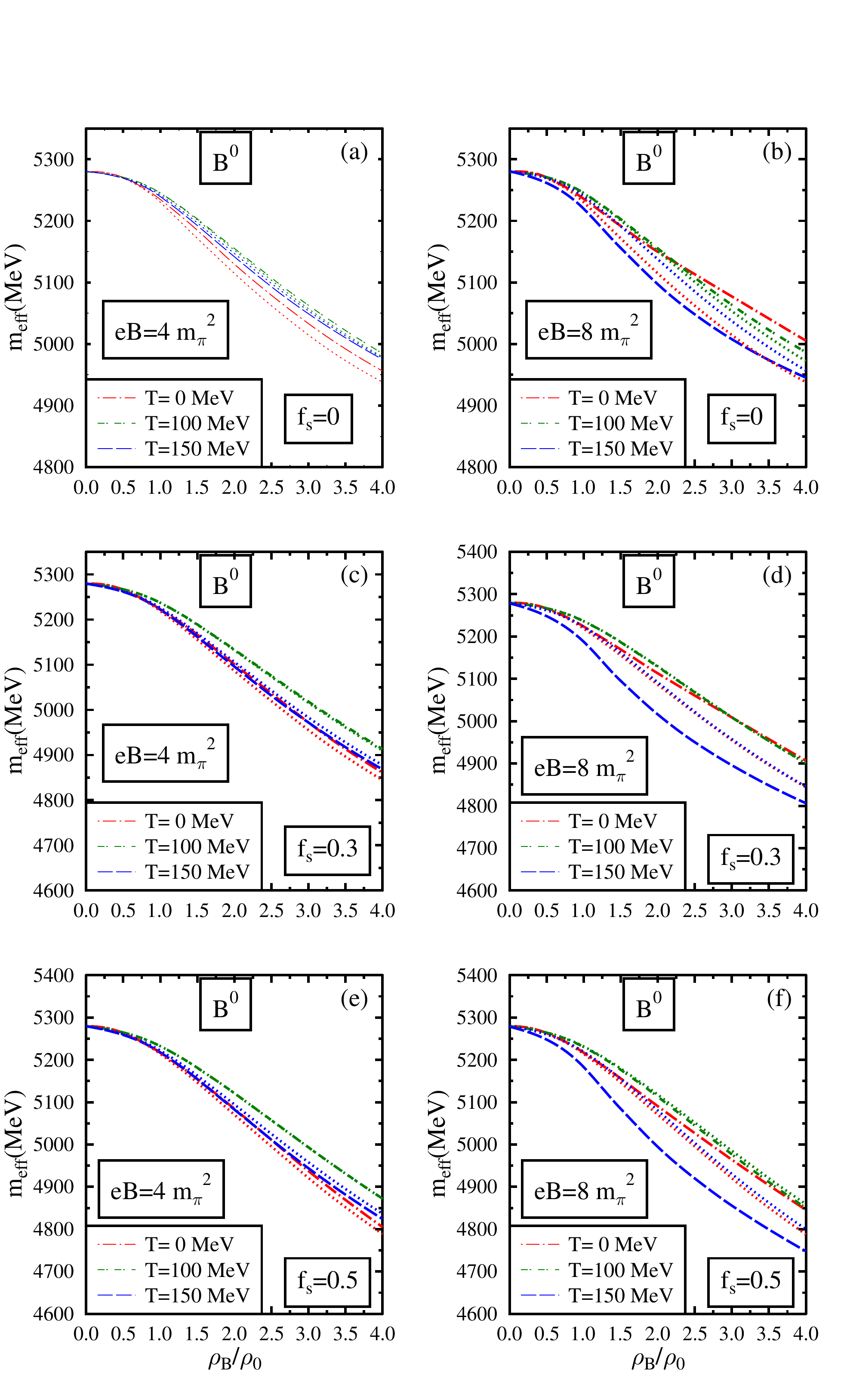}
\caption{The effective mass of $B^0$ meson in isospin asymmetric ($\eta$=0.5) hadronic matter is plotted as a function of the baryon density $\rho_{B} $/$\rho_{0}$  for different values of temperature T= 0, 100, and 150 MeV. These are plotted at magnetic fields $eB=4m_{\pi}^2$ and $eB=8m_{\pi}^2$, for a fixed value of strangeness fraction $f_s$ = 0, 0.3, 0.5. The effects of the anomalous magnetic moment of baryons are taken into account (dashed lines) and compared to the case when these effects are not considered (dotted lines).}
\label{mBzero}
\end{figure}

\begin{figure}[htbp]
\includegraphics[height=18.15cm, width=11cm, keepaspectratio=true]{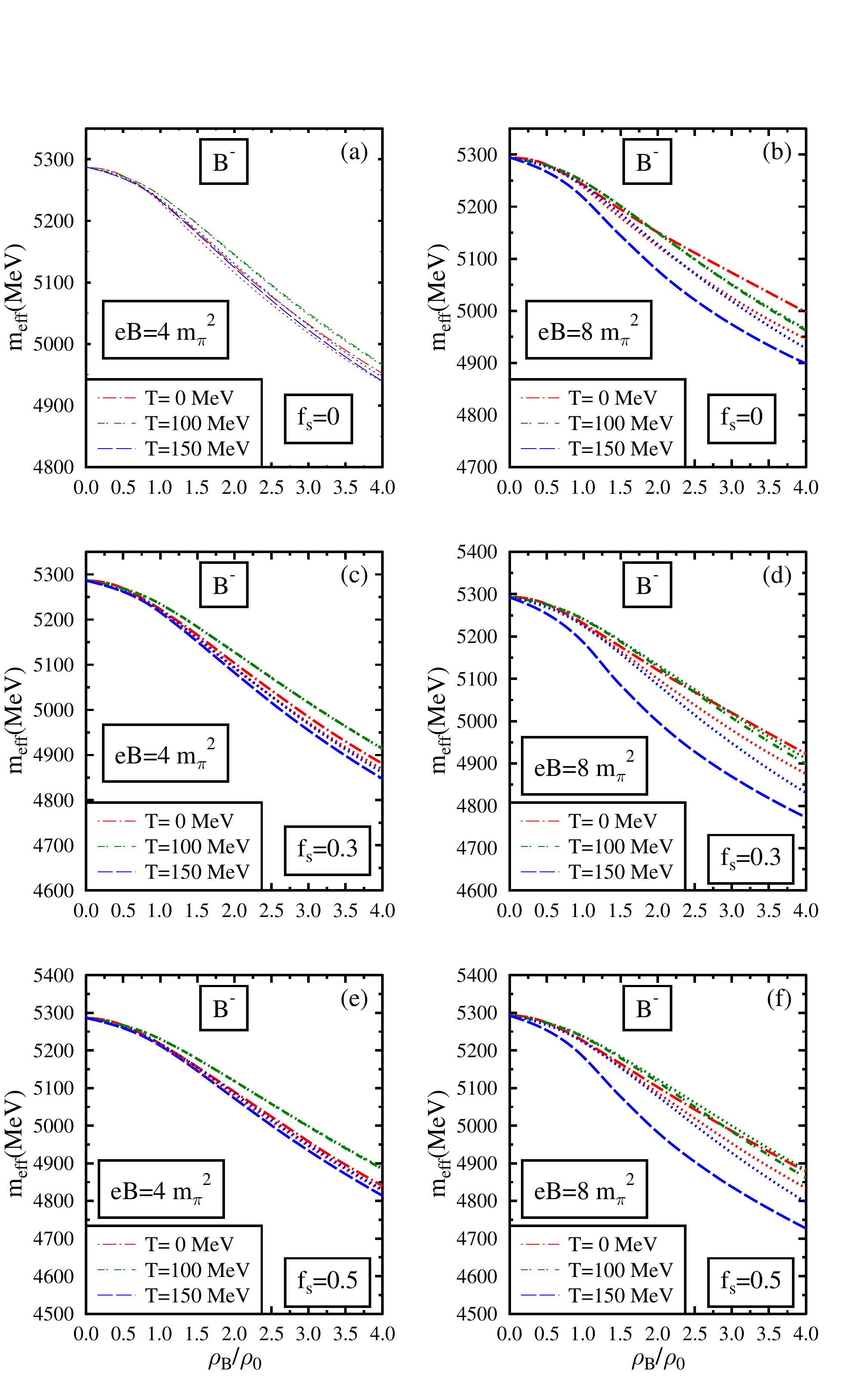}
\caption{The effective mass of $B^-$ meson in isospin asymmetric ($\eta$=0.5) hadronic matter is plotted as a function of the baryon density $\rho_{B} $/$\rho_{0}$  for different values of temperature T= 0, 100, and 150 MeV. These are plotted at magnetic fields $eB=4m_{\pi}^2$ and $eB=8m_{\pi}^2$, for a fixed value of strangeness fraction $f_s$ = 0, 0.3, 0.5. The effects of the anomalous magnetic moment of baryons are taken into account (dashed lines) and compared to the case when these effects are not considered (dotted lines).}
\label{mBminus}
\end{figure}

\begin{figure}[htbp]
\includegraphics[height=18.15cm, width=11cm, keepaspectratio=true]{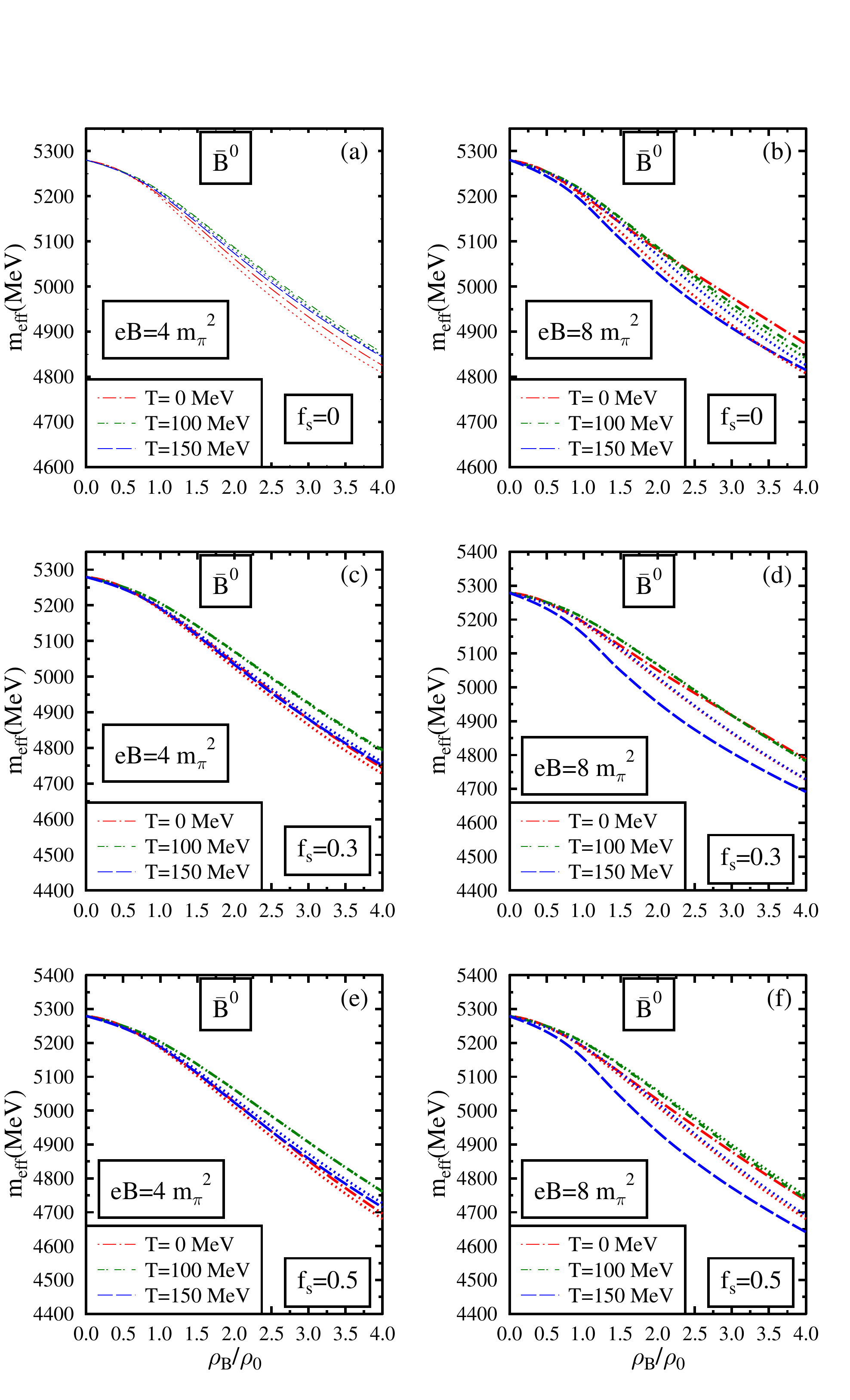}
\caption{The effective mass of $\bar{B^0}$ meson in isospin asymmetric ($\eta$=0.5) hadronic matter is plotted as a function of the baryon density $\rho_{B} $/$\rho_{0}$  for different values of temperature T= 0, 100, and 150 MeV. These are plotted at magnetic fields $eB=4m_{\pi}^2$ and $eB=8m_{\pi}^2$, for a fixed value of strangeness fraction $f_s$ = 0, 0.3, 0.5. The effects of the anomalous magnetic moment of baryons are taken into account (dashed lines) and compared to the case when these effects are not considered (dotted lines).}
\label{mBzerobar}
\end{figure}

\begin{figure}[htbp]
\includegraphics[height=18.15cm, width=11cm, keepaspectratio=true]{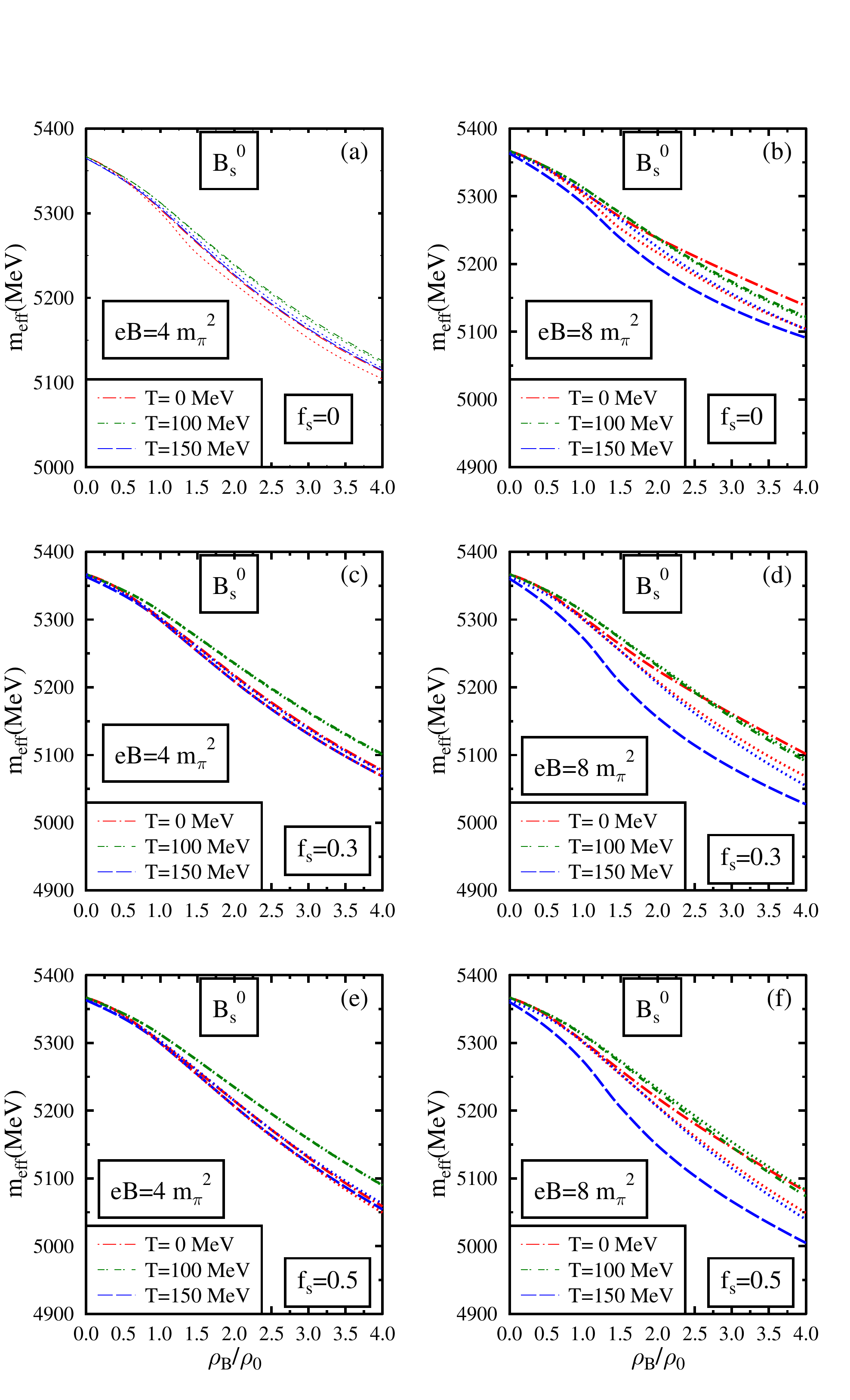}
\caption{The effective mass of ${B_s}^0$ meson in isospin asymmetric ($\eta$=0.5) hadronic matter is plotted as a function of the baryon density $\rho_{B} $/$\rho_{0}$  for different values of temperature T= 0, 100, and 150 MeV. These are plotted at magnetic fields $eB=4m_{\pi}^2$ and $eB=8m_{\pi}^2$, for a fixed value of strangeness fraction $f_s$ = 0, 0.3, 0.5. The effects of the anomalous magnetic moment of baryons are taken into account (dashed lines) and compared to the case when these effects are not considered (dotted lines).}
\label{mBszero}
\end{figure}

\begin{figure}[htbp]
\includegraphics[height=18.15cm, width=11cm, keepaspectratio=true]{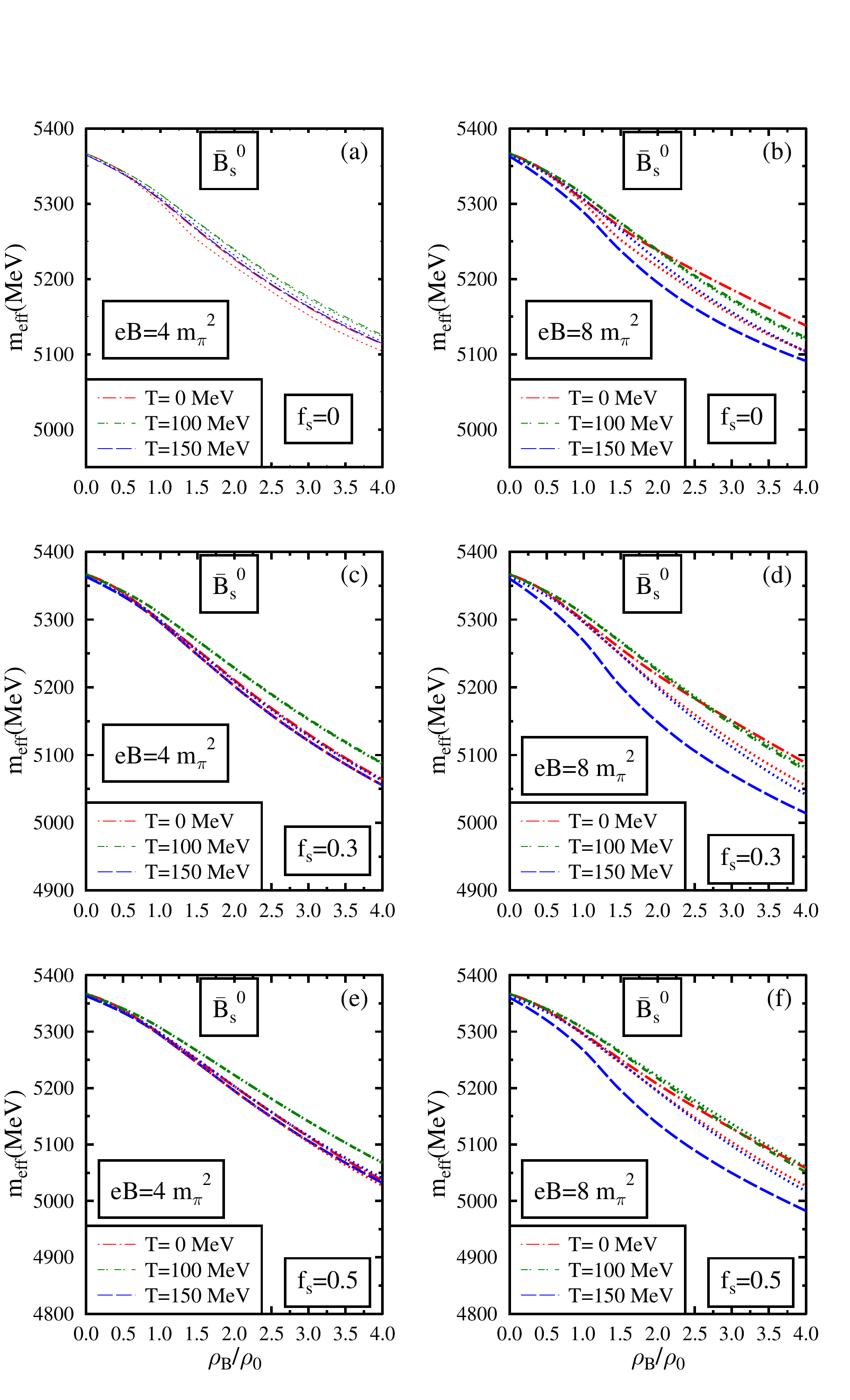}
\caption{The effective mass of $\bar{B_s}^0$ meson in isospin asymmetric ($\eta$=0.5) hadronic matter is plotted as a function of the baryon density $\rho_{B} $/$\rho_{0}$  for different values of temperature T= 0, 100, and 150 MeV. These are plotted at magnetic fields $eB=4m_{\pi}^2$ and $eB=8m_{\pi}^2$, for a fixed value of strangeness fraction $f_s$ = 0, 0.3, 0.5. The effects of the anomalous magnetic moment of baryons are taken into account (dashed lines) and compared to the case when these effects are not considered (dotted lines).}
\label{mBszerobar}
\end{figure}

In figures \ref{mBplus}, \ref{mBzero}, \ref{mBminus}, \ref{mBzerobar}, \ref{mBszero}, and \ref{mBszerobar}, the masses of the open bottom mesons $B^+$, $B^0$, $B^-$, $\bar{B^0}$, ${B_s}^0$, and $\bar{B_s}^0$ in the isospin asymmetric hadronic matter ( $\eta$=0.5)  are plotted as functions of baryon density $\rho_{B} $/$\rho_{0}$ for T=0 MeV, 100 MeV, and 150 MeV. Each subplot corresponds to a fixed value of magnetic fields eB= $4{m_{\pi}}^2$, $8{m_{\pi}}^2$ and  strangeness fraction $f_s$=0, 0.3, and 0.5. The in-medium masses of the open bottom mesons are calculated by solving the dispersion relations given in eq.(\ref{disp_relation}), with their self-energies being given by eq.(\ref{selfenergy_BplusB0}), eq.(\ref{selfenergy_BminusBbar0}), and eq.(\ref{selfenergy_Bs)}). The mass modifications of these mesons in the magnetized medium are due to the Weinberg-Tomozawa term, the scalar meson exchange term, and the range terms. The magnetic field contributes an additional positive modification to the masses of charged $B^\pm$ mesons through Landau quantization in accord with eq.(\ref{charged_effective_mass}). In this investigation, the B meson decay constant is be taken to be $f_B$=190.6 MeV\cite{AM_Divakar_PRC_2015} and their vacuum masses are $m_{B^+}$ = $m_{B^-}$ = 5279.32 MeV and $m_{B^0}$ = $m_{\bar{B^0}}$ = 5279.63 MeV. The decay constant of $B_s$ meson is taken to be $f_{B_s}$= 224 MeV \cite{Divakar_AM_IntJMod_2014}, and their vacuum masses are taken to be $m_{{B_s}^0}$ = $m_{\bar{B_s}^0}$ = 5366.89 MeV. 

The isospin symmetric part of the Weinberg-Tomozawa term is repulsive for $B$ mesons and attractive for $\bar{B}$ mesons. Hence in the symmetric hadronic matter, the Weinberg-Tomozawa term causes an increase in the masses of $B^+$, $B^0$ and leads to a mass drop of $B^-$ and $\bar{B^0}$. The isospin asymmetric part of the Weinberg-Tomozawa term is more repulsive for $B^0$ meson as compared to $B^+$. The same term is more attractive for the $\bar{B^0}$ meson as compared to the $B^-$ meson. Hence in the asymmetric matter, this contribution leads to a larger positive mass shift mass for the $B^0$ meson compared to $B^+$ and a more significant mass drop for the $\bar{B^0}$  meson as compared to the $B^-$ meson. The scalar meson exchange term is attractive, whereas the first range term is repulsive for $B$ and $\bar{B}$ mesons. In the asymmetric matter($\eta=0.5$), $\delta$ is negative. The $\delta$ term part of the scalar-meson exchange term results in a larger mass drop for $B^0$ as compared to $B^+$ and a more significant mass drop for $\bar{B^0}$ meson as compared to $B^-$ meson. Similarly, in the asymmetric medium, the $\delta$ term makes the first range term more repulsive for $B^0$ compared to $B^+$ meson and more repulsive for $\bar{B^0}$ compared to $B^-$ meson. The $d_1$ and $d_2$ range terms whose magnitude depends on the scalar densities of baryons are attractive for all mesons.

The in-medium masses of the $B$ and $\bar{B}$ mesons drop with an increase in the baryon density due to the attractive contribution from the scalar meson exchange term, $d_1$, and $d_2$ range terms. The slow rate of mass drop for $B$ and $\bar{B}$ at small densities is due to the repulsive first range term, which dominates over the $d_1$ and $d_2$ range terms in this regime. However, at higher densities, the $d_1$ and $d_2$ range terms become dominant since the scalar densities increase monotonically with an increase in number density compared to the sub-linear variation of $\sigma$. Since the magnitude of the Weinberg-Tomozowa term increases with baryon density, it leads to a large attractive contribution to the $\bar{B}$ mesons. However, for $B$ mesons, the Weinberg-Tomozowa term is repulsive and opposes the effect of the attractive contributions from other terms. At temperature T=0 MeV and $f_s$=0, under eB= $\ 4m_\pi^2$, incorporating the effects of AMM, $B^+$ and $B^0$ experience a mass drop of approximately 29.06 MeV and 44.80 MeV at $\rho_B$= 1$\rho_0$. Under similar conditions, $B^-$ and $\bar{B^0}$ experience a mass drop of 45.83 MeV and 78.72 MeV. At $\rho_B$= 4$\rho_0$, the mass drop of  $B^+$ and $B^0$ mesons increases to  261.32 MeV and 323.40 MeV, respectively, and that of  $B^-$  and $\bar{B^0}$ mesons increases to 327.10 MeV and 454.44 MeV.

As the value of $f_s$ increases, the in-medium mass of $B$ and $\bar{B}$ mesons drops further. At T=0 MeV under a magnetic field of eB= $\ 4m_\pi^2$, incorporating the effects of AMM, the mass drop of $B^+$, $B^0$, $B^-$, and $\bar{B^0}$ mesons in MeV are 42.40, 56.80, 56.60 and 88.12, respectively, at $\rho_B$= 1$\rho_0$ when $f_s$=0.3. These mass drops are larger than the corresponding mass drops at $f_s$=0 due to the increase in the cumulative scalar densities of baryons with an increase in $f_s$. Hence the $d_1$ and $d_2$ range terms are more attractive in the strange hadronic matter compared to the nuclear matter. Moreover, the scalar meson exchange term and the first range term rely on the fluctuations of the scalar fields  $\sigma'$ and $\delta'$, whose values are dependent on $f_s$. The magnitude of the Weinberg-Tomozawa term and its contribution to the mass of $B$ and $\bar{B}$ becomes weaker in the presence of hyperons \cite{AM_Divakar_PRC_2015}. As a result, the Weinberg-Tomozawa term becomes less repulsive for $B$ meson and less attractive for $\bar{B}$ in the strange hadronic medium compared to the nuclear medium. This effect of $f_s$ is amplified at high densities. Moreover, the weakening of the Weinberg-Tomozawa term in the strange hadronic medium reduces the mass asymmetry of the mesons belonging to isospin doublets.

At eB= $4{m_{\pi}}^2$, when the effects of AMM are taken into account, the in-medium masses of $B^+$, $B^0$, $B^-$, $\bar{B^0}$ mesons are observed to increase (mass drop decreases) as the temperature is increased from T=0 MeV to T=100 MeV. At T=0 MeV, for $f_s$=0.3 the mass drop of  $B^+$, $B^0$, $B^-$ and $\bar{B^0}$ mesons in MeV are 344.45, 418.90, 398.63, and 536.40, respectively, at $\rho_B$= 4$\rho_0$. Under similar conditions, when the temperature is increased to T=100 MeV, the corresponding mass drop reduces to 310.62, 366.43, 366.10, and 485.63 MeV. The above behavior is due to the increase in the magnitude of $\sigma$ with an increase in temperature( till the turnaround temperature of T=110 MeV at $\rho_B$ = $\rho_0$), which reduces the attractive contribution of the scalar meson exchange term. As the cumulative scalar density of baryons undergoes a drop with temperature till T=110 MeV under these conditions, the magnitude of the attractive $d_1$ and $d_2$  range terms also weakens. 

At eB= $4{m_{\pi}}^2$, as the temperature is further increased above the turnaround temperature( T=110 MeV at $\rho_B$ = $\rho_0$), the in-medium masses begin to drop (mass drop increase) since the value of scalar field $\sigma$ significantly decreases with an increase in temperature. Although the magnitude of $\delta$ decreases with an increase in temperature, reducing the asymmetric contribution,  this variation is subdominant compared to the variation of $\sigma$. Moreover, the scalar density of protons rises at high temperatures, contrary to the scalar densities of neutrons which decrease with temperature in the presence of the magnetic field. At T=150 MeV, due to the cumulative contributions of all these terms, there is an increase in the magnitude of the net attractive contributions, and we observe a larger mass drop for $B$ and $\bar{B}$ mesons compared to T=100 MeV. At T=150 MeV, for $f_s$=0.3, under a magnetic field of eB= $\ 4m_\pi^2$ and $\rho_B$= 4$\rho_0$. incorporating the effects of AMM, the mass drops of  $B^+$, $B^0$, $B^-$, and $\bar{B^0}$ mesons in MeV are observed to be 377.91, 411.43, 431.93, and 528.70, respectively.

For eB= $8{m_{\pi}}^2$ and $f_s=0$, the mass of these mesons at T=0 MeV is smaller compared to their mass at T=100 MeV till $\rho_B=1.7\rho_0$. For $f_s=0.3$, a similar behaviour is observed till $\rho_B=2\rho_0$. Above this density, the masses of these mesons at T=0 MeV are larger than their masses at T=100 MeV, owing to the reduction of the turnaround temperature for $\sigma$  with an increase in the magnetic field. Moreover, the effect of temperature on these mesons is more significant at larger magnetic fields. 

At T=0 MeV, the in-medium mass of $B$ and $\bar{B}$ mesons increases with an increase in the magnetic field. The magnetic field introduces modifications of $B$ and $\bar{B}$ mesons due to the modification of the scalar fields and the scalar densities of the baryons. This mass modification induced by the magnetic field is significant at large densities and small values of $f_s$. In addition, the charged $B^+$ and $B^-$ mesons undergo further positive modification due to the landau quantization in accord with eq.(\ref{charged_effective_mass}). At small baryon densities, $B^+$ and $B^-$ undergo a larger positive mass shift from the magnetic field due to this Landau quantization effect which is absent in the case of neutral $B$ mesons. At  T=0 MeV, when AMM effects are incorporated, the magnitude of the cumulative scalar density of baryons drops with the magnetic field. This reduction of scalar density weakens the attractive contributions from the range terms, and hence the masses of  $B^0$ and $\bar{B^0}$ undergo a  small positive mass modification by the magnetic field at T=0 MeV. With an increase in the magnetic field, at large baryon densities, the attractive contribution of the $d_2$ range term weakens more for $B^0$ and $\bar{B^0}$ mesons compared to $B^+$ and $B^-$ mesons. Hence at large densities, $B^0$ and $\bar{B^0}$ experience larger positive mass shifts due to the modification of scalar fields and scalar densities induced by an increase in the magnetic field. At T=0 MeV, for $f_s=0.3$, incorporating the effects of AMM, as the magnetic field is increased from eB= $\ 4m_\pi^2$ to eB= $\ 8m_\pi^2$, $B^0$ and $\bar{B^0}$ experience an increase of mass of  approximately 0.71 (45.88) MeV and 0.72 (44.41) MeV, respectively, at $\rho_B$= 1$\rho_0$ (4$\rho_0$). Under similar conditions, $B^+$ and $B^-$ experience a mass increase of  7.91 (42.63) MeV and 7.94 (42.3) MeV, respectively, at $\rho_B$= 4$\rho_0$. 

 For $\eta=0.5$, $f_s=0$,  the hadronic matter is composed of only neutrons. In this case, when the AMM effects are ignored, the neutral $B^0$ and $\bar{B^0}$ mesons experience no modifications due to the change in the magnetic field at T=0 MeV. When AMM effects are ignored, neutrons being electrically neutral are not subject to modification by the magnetic field. A detailed discussion of the medium modifications of $B$ and $\bar{B}$ mesons in the magnetized nuclear matter at T=0 MeV and a comparison of their in-medium behavior with that of $D$ and $\bar{D}$ mesons using the chiral effective model is given in Ref \cite{Dhale}. For strange hadronic matter, at T=0 MeV, without incorporating AMM effects, the masses of $B^0$ and $\bar{B^0}$ mesons are observed to marginally drop with an increase in the magnetic field owing to the marginal drop in the magnitude of $\sigma$. At temperature T=0 MeV the in-medium masses of  $B$ and $\bar{B}$ are observed to be smaller when AMM effects are not taken into account as compared to when these effects are incorporated. 

As the temperature is increased from T=0 MeV to T=90 MeV, the magnitude of the positive mass shift due to the modification of the scalar fields and scalar densities induced by the magnetic field begins to decline. At T $>$ 90 MeV, when AMM effects are incorporated, the functional dependence of the masses of $B^0$ and $\bar{B^0}$ mesons on the magnetic field changes direction. In this temperature regime, the in-medium masses of these mesons instead experience a drop with an increase in the magnetic field. Such a behavior can be attributed to the different nature of the modification of scalar fields and cumulative scalar densities due to the magnetic field at higher temperatures. For $B^+$ and $B^-$ mesons,  there is an interplay between the positive mass shift from the Landau quantization effect and the negative mass shift due to the modification of scalar fields due to the magnetic field when T=90-110 MeV. For T=100 MeV and $f_s=0$, when AMM effects are incorporated, the effect of Landau quantization becomes dominant till $\rho_B=2.9 \rho_0$, and above this density, the charged mesons undergo a drop with an increase in the magnetic field. For $f_s=0.5$, at the same temperature value, this mass drop induced by the magnetic field is observed above $\rho_B=1.9 \rho_0$. 

At T=150 MeV, the amount of mass drop experienced by both $B$ and $\bar{B}$ mesons as a function of the magnetic field is more than that at T=100 MeV. Moreover, the mass drop of these mesons induced by the magnetic field at high temperature is larger for larger values of $f_s$. At T=150 MeV, for $f_s=0.3$, incorporating the effects of AMM, when the magnetic field is increased from eB= $\ 4m_\pi^2$ to eB= $\ 8m_\pi^2$, $B^0$ and $\bar{B^0}$ experience a mass drop of 62 MeV and 59.7 MeV,respectively, at $\rho_B$= 4$\rho_0$. Under similar conditions, $B^+$ and $B^-$ mesons experience a further mass drop of 75.7 MeV and 74.5 MeV, respectively, at $\rho_B$= 4$\rho_0$. Despite the positive modification from the Landau quantization effect, the charged $B$ mesons undergo a mass drop due to the magnetic field at high temperature and large baryon density. This behavior is because the $d_2$ range term becomes more attractive for these mesons under these conditions. At T=150 MeV, the in-medium masses of  $B$ and $\bar{B}$ are observed to be larger when AMM effects are neglected as compared to the case when these effects are incorporated.

The masses of ${B_s}^0$ and $\bar{B_s}^0$ also drop with an increase in baryon density, although the magnitude of the drop is smaller compared to $B$ and $\bar{B}$ mesons. For nuclear matter, this mass drop is dominantly due to the increase in scalar density of nucleons with an increase in baryon density leading to large attractive contributions from the $d_1$ range term. The magnitude of the attractive scalar meson exchange is also significant, but its contribution gets saturated at large densities owing to the saturation behavior of $\zeta$ in the nuclear matter. The Weinberg-Tomozowa term and the $d_2$ range term for ${B_s}$ mesons are dependent only on the number densities and scalar densities of hyperons, respectively. Hence the contribution of these terms vanishes in the nuclear matter. Due to the absence of the Weinberg-Tomozowa term, ${B_s}^0$ and $\bar{B_s}^0$ exhibit mass degeneracy in the nuclear matter. Such a degeneracy is also observed for strange-charmed $D_s$ mesons in the nuclear matter \cite{Divakar_AM_AdvHighEner_2015,Magstrange}. At $f_s$=0, T=0 MeV , under eB= $\ 4m_\pi^2$, incorporating the effects of AMM, both $B_s$ mesons experience a mass drop of 62.35 MeV at $\rho_B$= 1$\rho_0$. This mass drop increases to 253.54 MeV at $\rho_B$= 4$\rho_0$. The coefficient of the attractive scalar meson exchange term for ${B_s}$ mesons ($m^2_{B_S}/\sqrt{2}f_{B_s}$) is larger than that of ${D_s}$ mesons ($m^2_{D_S}/\sqrt{2}f_{D_s}$). Hence at T=0 MeV, the mass drop of ${B_s}$ mesons in the magnetized strange hadronic medium is larger than that of ${D_s}$ mesons given in Ref \cite{Magstrange}. 

With an increase in  $f_s$, the hadronic matter gets populated with hyperons. In this case, the Weinberg-Tomozowa term and the $d_2$ range term contribute. The Weinberg-Tomozowa term being attractive for $\bar{B_s}^0$ meson and repulsive for ${B_s}^0$ meson ensures a more significant mass drop for $\bar{B_s}^0$ in magnetized strange hadronic medium and breaks their mass degeneracy. The $d_2$ range term contributes equally to both mesons. The magnitude of the attractive $d_1$ range term is larger in the strange hadronic medium than in the nuclear medium. Even the magnitude of the attractive scalar meson exchange term increases with an increase in $f_s$ since $\zeta$ drops significantly, especially at large densities. Hence at large baryon densities, ${B_s}^0$ and $\bar{B_s}^0$ mesons experience a larger mass drop in the strange hadronic medium as compared to the nuclear medium. 

The temperature dependence of the in-medium mass of ${B_s}^0$ and $\bar{B_s}^0$ mesons is qualitatively similar to that of $B$ and $\bar{B}$ mesons. However, due to the absence of number densities of nucleons in the Weinberg-Tomozowa term and the absence of the scalar densities of nucleons in the $d_2$ range term, the overall temperature dependence is weaker for ${B_s}$ mesons. Moreover, the $\zeta$ field, which contributes to the scalar meson exchange term and the first range term for $B_s$ mesons, undergoes marginal modifications due to temperature compared to that of $\sigma$, which contributes to the $B$ and $\bar{B}$ mesons. For $f_s$= 0.3, at T=0 MeV and , under eB= $4m_\pi^2$, incorporating the effects of AMM, ${B_s}^0$ and $\bar{B_s}^0$ experience a mass drop of 289.61 MeV and 302.90 MeV, respectively, at $\rho_B$= 4$\rho_0$. Under similar conditions, when the temperature is increased to T=100 MeV, their mass drop reduces to 265.69 MeV and 279 MeV. At T=150 MeV, the corresponding mass drop increases to 298.19 MeV and 311.49 MeV, respectively, compared to the mass drop at T=100 MeV.

The mass modification of bottom-strange mesons as a function of the magnetic field arises from the variation of the scalar field $\zeta$ and the scalar densities of baryons due to the magnetic field. For $f_s=0$, when AMM effects are taken into account, the in-medium masses of bottom-strange mesons increase with an increase in the magnetic field when the temperature is below T= 90 MeV. For $f_s=0.3$, at T=0 MeV, as the magnetic field is increased from eB= $\ 4m_\pi^2$ to eB= $\ 8m_\pi^2$, the in-medium masses of ${B_s}^0$ and $\bar{B_s}^0$ mesons increase by approximately 24.29 MeV and 24.16 MeV, respectively, at $\rho_B$= 4$\rho_0$.  Above T=90 MeV, the in-medium masses of ${B_s}$ mesons begin to drop with an increase in the magnetic field. For $f_s=0.3$, at T=150 MeV incorporating the effects of AMM, as the magnetic field is increased from eB= $\ 4m_\pi^2$ to eB= $\ 8m_\pi^2$, ${B_s}^0$ and $\bar{B_s}^0$  experience a decrease of mass of  approximately 41.50 MeV and 41.30 MeV, respectively, at $\rho_B$= 4$\rho_0$. The effect of the magnetic field on ${B_s}$ mesons is smaller when the effects of AMM are neglected.

\begin{figure}[htbp]
\includegraphics[height=18.15cm, width=11cm, keepaspectratio=true]{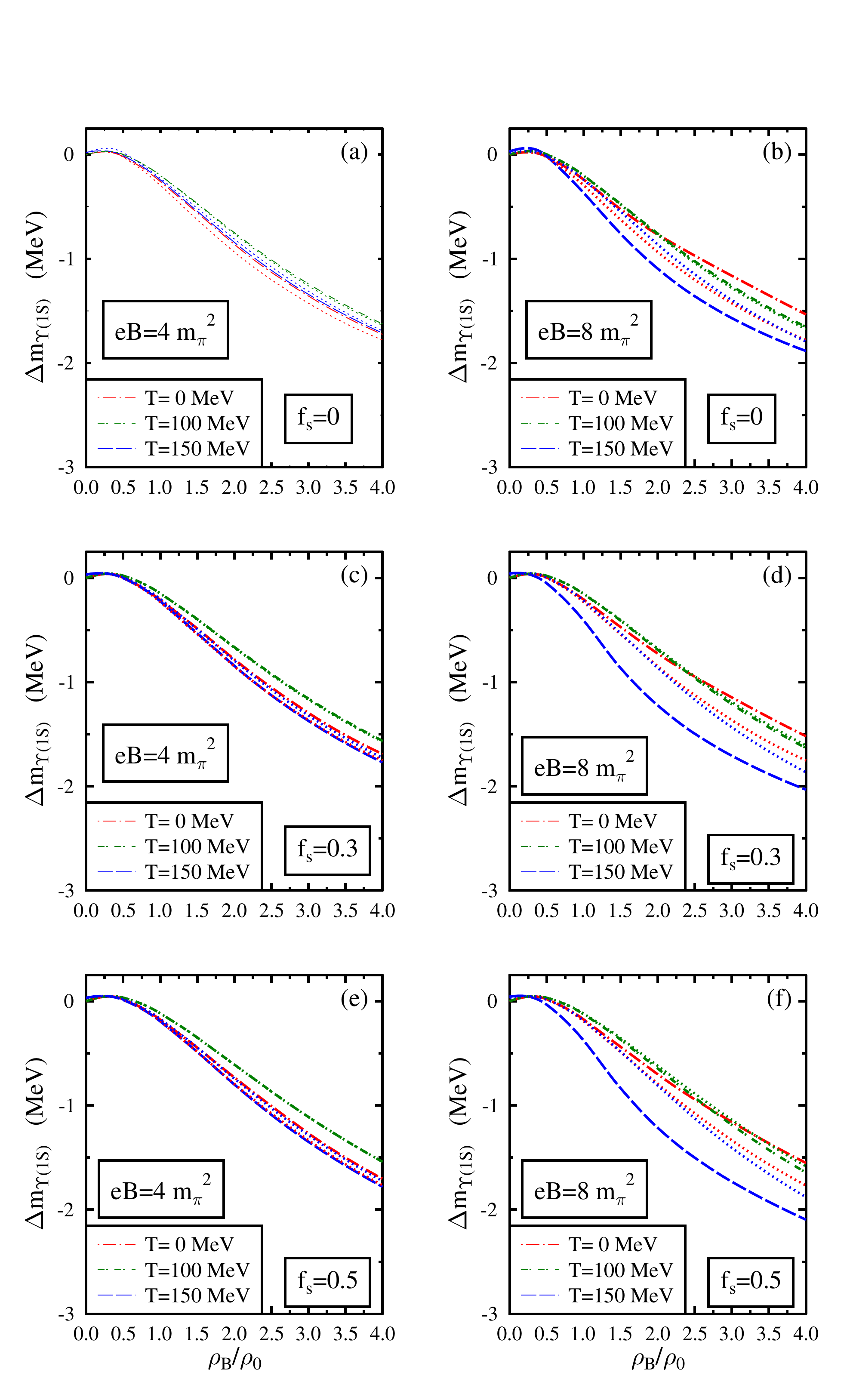}

\caption{The mass shift of $\Upsilon(1S)$ in isospin asymmetric ($\eta$=0.5) hadronic matter is plotted as a function of the baryon density $\rho_{B} $/$\rho_{0}$  for different values of temperature T= 0, 100, and 150 MeV. These are plotted at magnetic fields $eB=4m_{\pi}^2$ and $eB=8m_{\pi}^2$, for a fixed value of strangeness fraction $f_s$ = 0, 0.3, 0.5. The effects of the anomalous magnetic moment of baryons are taken into account (dashed lines) and compared to the case when these effects are not considered (dotted lines).}
\label{mUpsilon1S}
\end{figure}

\begin{figure}[htbp]
\includegraphics[height=18.15cm, width=11cm, keepaspectratio=true]{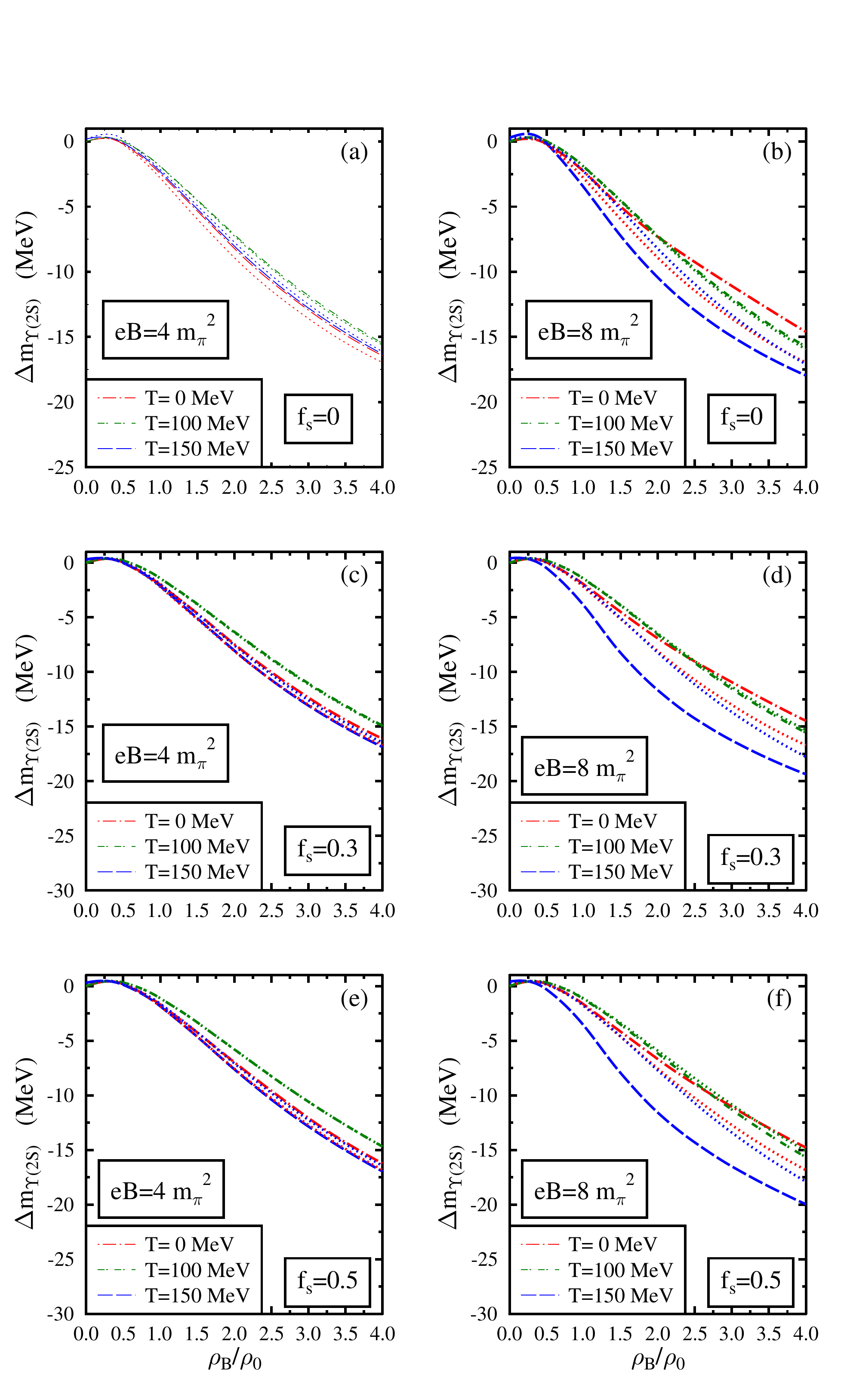}
\caption{The mass shift of $\Upsilon(2S)$ in isospin asymmetric ($\eta$=0.5) hadronic matter is plotted as a function of the baryon density $\rho_{B} $/$\rho_{0}$  for different values of temperature T= 0, 100, and 150 MeV. These are plotted at magnetic fields $eB=4m_{\pi}^2$ and $eB=8m_{\pi}^2$, for a fixed value of strangeness fraction $f_s$ = 0, 0.3, 0.5. The effects of the anomalous magnetic moment of baryons are taken into account (dashed lines) and compared to the case when these effects are not considered (dotted lines).}
\label{mUpsilon2S}
\end{figure}

\begin{figure}[htbp]
\includegraphics[height=18.15cm, width=11cm, keepaspectratio=true]{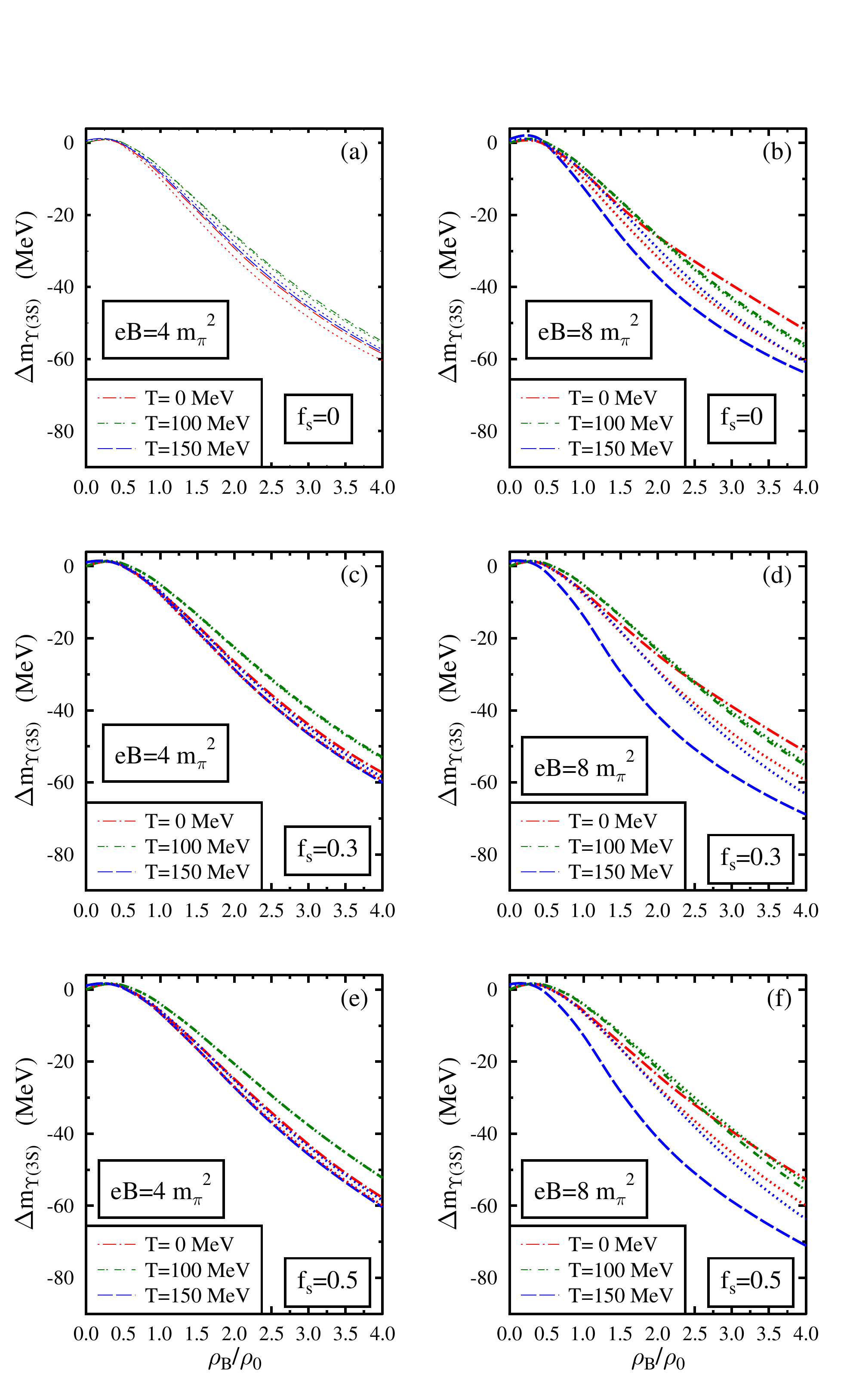}
\caption{The mass shift of $\Upsilon(3S)$ in isospin asymmetric ($\eta$=0.5) hadronic matter is plotted as a function of the baryon density $\rho_{B} $/$\rho_{0}$  for different values of temperature T= 0, 100, and 150 MeV. These are plotted at magnetic fields $eB=4m_{\pi}^2$ and $eB=8m_{\pi}^2$, for a fixed value of strangeness fraction $f_s$ = 0, 0.3, 0.5. The effects of the anomalous magnetic moment of baryons are taken into account (dashed lines) and compared to the case when these effects are not considered (dotted lines).}
\label{mUpsilon3S}
\end{figure}

\begin{figure}[htbp]
\includegraphics[height=18.15cm, width=11cm, keepaspectratio=true]{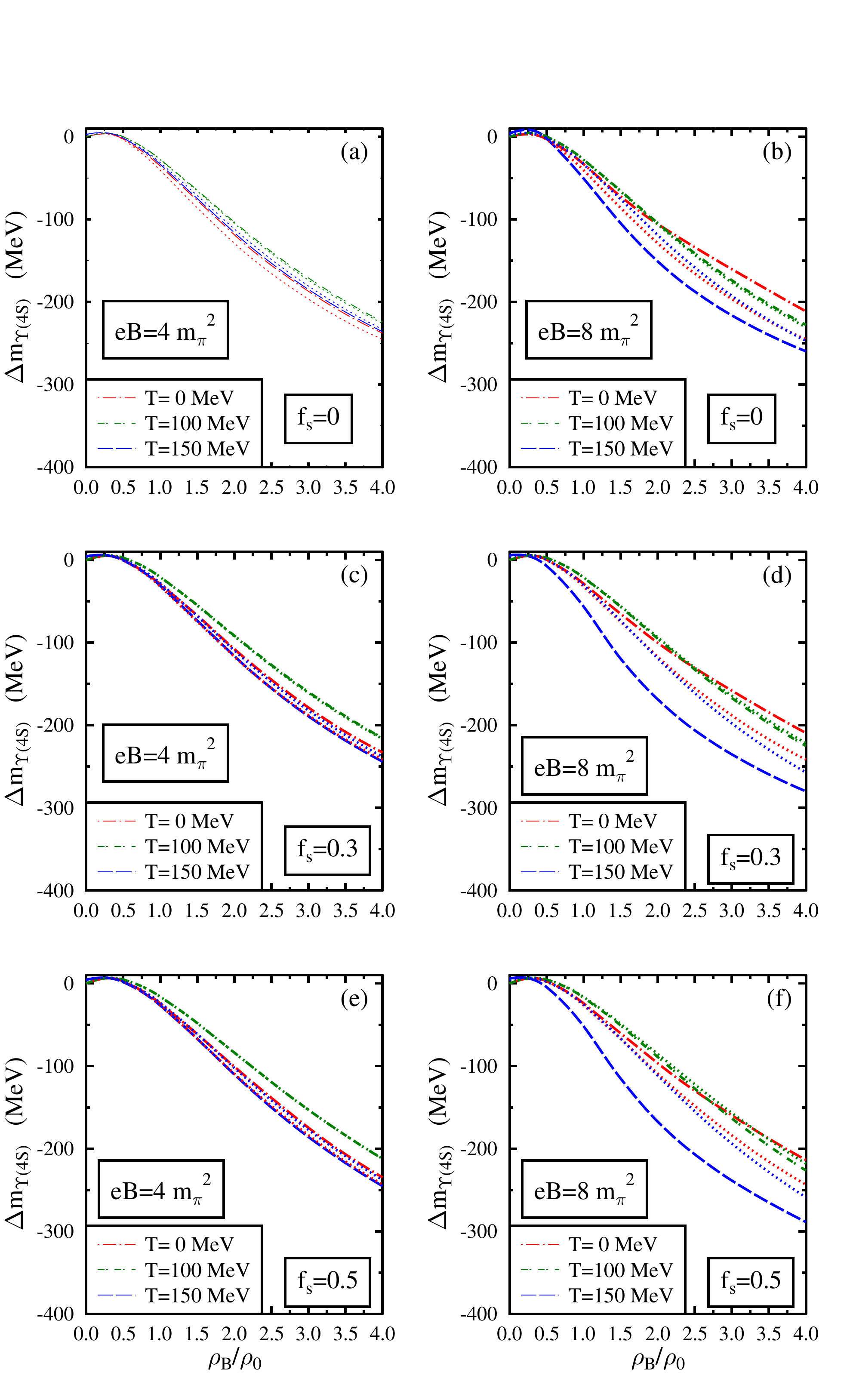}
\caption{The mass shift of $\Upsilon(4S)$ in isospin asymmetric ($\eta$=0.5) hadronic matter is plotted as a function of the baryon density $\rho_{B} $/$\rho_{0}$  for different values of temperature T= 0, 100, and 150 MeV. These are plotted at magnetic fields $eB=4m_{\pi}^2$ and $eB=8m_{\pi}^2$, for a fixed value of strangeness fraction $f_s$ = 0, 0.3, 0.5. The effects of the anomalous magnetic moment of baryons are taken into account (dashed lines) and compared to the case when these effects are not considered (dotted lines).}
\label{mUpsilon4S}
\end{figure}

\begin{figure}[htbp]
\includegraphics[height=18.15cm, width=11cm, keepaspectratio=true]{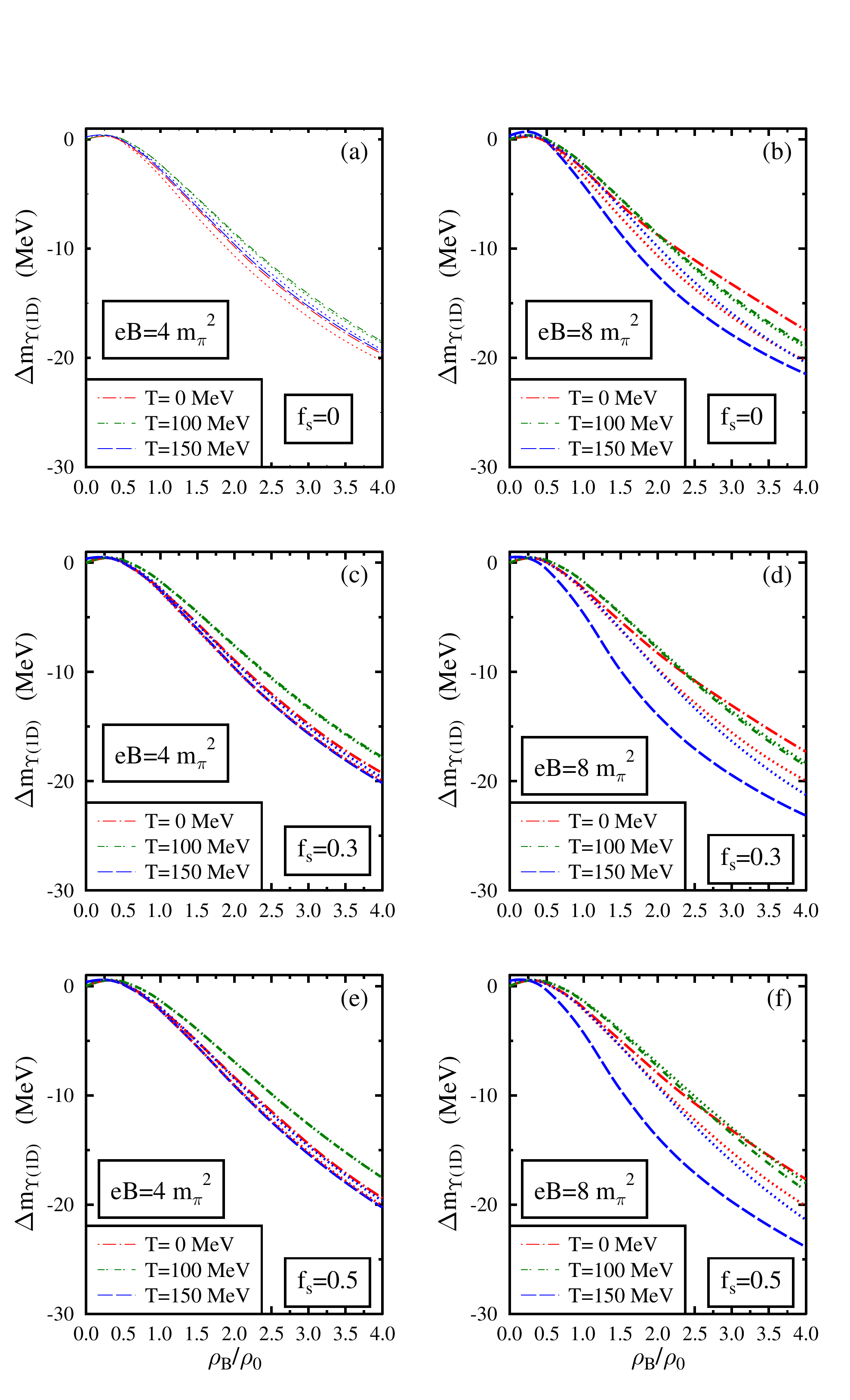}
\caption{The mass shift of $\Upsilon(1D)$ in isospin asymmetric ($\eta$=0.5) hadronic matter is plotted as a function of the baryon density $\rho_{B} $/$\rho_{0}$  for different values of temperature T= 0, 100, and 150 MeV. These are plotted at magnetic fields $eB=4m_{\pi}^2$ and $eB=8m_{\pi}^2$, for a fixed value of strangeness fraction $f_s$ = 0, 0.3, 0.5. The effects of the anomalous magnetic moment of baryons are taken into account (dashed lines) and compared to the case when these effects are not considered (dotted lines).} 
\label{mUpsilon1D}
\end{figure}

The mass shifts of the upsilon states in the magnetized strange medium are calculated from the medium change of the fourth power of the dilaton field  $\chi^4-{\chi_0}^4$ as well as the terms proportional to $\sigma ' (=\sigma-\sigma_0)$, $\zeta ' (=\zeta-\zeta_0)$ in accord with eq.(\ref{massupsilon}) and eq.(\ref{massivegluoncondensate}). In figures \ref{mUpsilon1S}, \ref{mUpsilon2S}, \ref{mUpsilon3S}, \ref{mUpsilon4S}, and \ref{mUpsilon1D}, the masses of $\Upsilon(1S)$, $\Upsilon(2S)$, $\Upsilon(3S)$, $\Upsilon(4S)$, and $\Upsilon(1D)$ are plotted in isospin asymmetric hadronic medium ($\eta$=0.5) as functions of baryon density $\rho_{B} $/$\rho_{0}$ for T=0 MeV, 100 MeV, and 150 MeV at magnetic fields  eB= $4{m_{\pi}}^2$ and $8{m_{\pi}}^2$ for various values of strangeness fraction $f_s$. In this investigation, the mass of the bottom quark is taken to be $m_b$= 5.36 GeV. The vacuum masses of $\Upsilon(1S)$, $\Upsilon(2S)$, $\Upsilon(3S)$, $\Upsilon(4S)$ and $\Upsilon(1D)$ are 9460.3, 10023.26, 10355.2, 10579.4 and 10163.7 MeV, respectively. The values of the harmonic oscillator potential strength parameter, $\beta$ are calculated to be 1309.2, 915.4, 779.75 and 638.6 MeV for the $\Upsilon(1S)$, $\Upsilon(2S)$, $\Upsilon(3S)$, $\Upsilon(4S)$,respectively from their observed leptonic decay widths of 1.34, 0.612, 0.443 and 0.272 keV. The value of  $\beta$ for $\Upsilon (1D)$ is taken as 858 MeV, which is obtained through interpolation from mass versus $\beta$ plot for the upsilon states\cite{Amal2_upsilon}.

At small baryon densities (up to $\rho_{B}$= 0.3 - 0.5 $\rho_0$), upsilon states experience a marginal positive mass modification since the terms proportional to $\sigma '(=\sigma-\sigma_0)$ and $\zeta '(=\zeta-\zeta_0)$ in the eq.(\ref{massivegluoncondensate}) are positive and dominant at small densities. However, as baryon density increases, the term proportional to $\chi^4-{\chi_0}^4$ becomes dominant. As mentioned before, the magnitude of the dilaton field $\chi$ drops with an increase in baryon density. Since $\chi$ $<$ $\chi_0$ at finite density, the term $\chi^4-{\chi_0}^4$ is negative. This term results in a negative mass shift for upsilon states in the medium. The in-medium behavior of upsilon states is qualitatively governed by the $\chi^4-{\chi_0}^4$ term above $\rho_{B}$= 0.5$\rho_0$. Hence the in-medium masses of all upsilon states decrease with an increase in baryon density. All the upsilon states behave qualitatively similar in the magnetized hadronic medium. The magnitude of the mass shift for each state are different due to the difference in the magnitude of the integral (eq.(\ref{massupsilon})) calculated from their respective momentum wave functions. While the mass drop of $\Upsilon(1S)$ is marginal, the other excited states have a significant mass drop. It can be seen from the plots that $\Upsilon(4S)$ has the largest mass shift. For  $f_s$=0, at T=0 MeV , under eB= $\ 4m_\pi^2$, incorporating the effects of AMM, $\Upsilon(1S)$, $\Upsilon(2S)$, $\Upsilon(3S)$, $\Upsilon(4S)$, and $\Upsilon(1D)$ experience a mass shift (in MeV) of $-0.25 (-1.72)$, $-2.46(-16.44)$, $-8.76(-58.51)$, $-35.61(-237.79)$, and $-2.94(-19.65)$, respectively at $\rho_B$= 1$\rho_0$(4$\rho_0$). The magnitude of mass shifts of upsilon states decreases with an increase in  $f_s$ when T=0 MeV. For $f_s$=0.3, at T=0 MeV, under eB= $4m_\pi^2$, incorporating the effects of AMM, the mass shifts (in MeV) of upsilon states in the same order are  $-0.21(-1.69)$, $-2.01(-16.10)$, $-7.17(-57.31)$, $-29.15(-232.91)$ and $-2.41 (-19.25)$, respectively, at $\rho_B$= 1$\rho_0$(4$\rho_0$).

At $\rho_B= \rho_0$, for eB= $ 4m_\pi^2$,  when the effects of AMM are considered, the magnitude of mass shift of upsilon states initially decreases with an increase in temperature till T= 110 MeV. This behavior is due to the increase in the value of $\chi$ with an increase in the temperature in this regime. Hence the mass drop of upsilon states at T=100 MeV is smaller than that at T=0 MeV under eB= $4m_\pi^2$. In Ref \cite{Rajeshkumar2}, the mass shift of $\Upsilon(1S)$ in the hot magnetized nuclear matter calculated using QCD sum rules at T=100 MeV is also found to be smaller than the values at T=0 MeV. For $f_s$=0.3, at T=100 MeV under eB= $ 4m_\pi^2$, incorporating the effects of AMM, $\Upsilon(1S)$, $\Upsilon(2S)$, $\Upsilon(3S)$, $\Upsilon(4S)$, and $\Upsilon(1D)$ experience a mass shift (in MeV) of $-0.14(-1.56)$, $-1.41(-14.89)$, $-5.04(-53.01)$, $-20.51(-215.42)$, and $-1.69(-17.81)$, respectively, at $\rho_B$= 1$\rho_0$(4$\rho_0$). When eB= $4m_\pi^2$, above T=110 MeV, the value of $\chi$ begins to drop with a further increase in the temperature. Hence the mass shift of the upsilon states increases with temperature in this regime. This turnaround temperature decreases to below T=100 MeV with an increase in baryon density. Hence, the mass drop at T=150 MeV is larger than at T=100 MeV. At T=150 MeV, under similar medium conditions, the corresponding mass shifts (in MeV) of upsilon states in the same order are $-0.22(-1.77)$, $-2.14(-16.86)$, $-7.625(-60.01)$, $-30.97(-243.87)$, and $-2.56(-20.16)$, respectively, at $\rho_B$= 1$\rho_0$(4$\rho_0$). For $f_s=0.3$, when the magnitude of the magnetic field is increased to eB= $\ 8m_\pi^2$, the turnaround temperature of $\chi$ further lowers from T=110 MeV to T=60 MeV when the density is above $\rho_B=2.4\rho_0$. Hence at large magnetic fields and large baryon density, the magnitude of mass shift at T=100 MeV is larger than the mass shift at T=0 MeV. Moreover, the effect of temperature is pronounced at large magnetic fields, large baryon density, and large values of strangeness fraction, as can be seen from the plots. When the effects of AMM are ignored, the turnaround temperature does not vary significantly as a function of the magnetic field and baryon density.

At T=0 MeV, up to $\rho_B$= 1.5 $\rho_0$, since the scalar field modifications are weakly dependent on the magnetic field, the mass modifications of upsilon states are marginal as we increase the magnetic field. However, at higher densities, the effect of the magnetic field becomes significant on the mass shifts due to the larger value of $\chi$ due to the magnetic field. The upsilon states experience smaller mass shifts with an increase in the magnetic field at T=0 MeV. For  $f_s$=0.3, at T=0 MeV, under eB= $\ 8m_\pi^2$, incorporating the effects of AMM, the mass shifts (in MeV) of  $\Upsilon(1S)$, $\Upsilon(2S)$, $\Upsilon(3S)$, $\Upsilon(4S)$ and $\Upsilon(1D)$ are  $-0.20(-1.51)$, $-1.97(-14.47)$, $-7.03(-51.49)$, $-28.58(-209.25)$, and $-2.36(-17.30)$, respectively, at $\rho_B$= 1$\rho_0$(4$\rho_0$).These mass shifts are smaller than the values at eB= $4m_\pi^2$. When AMM effects of baryons are neglected, the effect of the magnetic field on the mass shift of upsilon states in the strange hadronic medium is marginal. At T=0 MeV, when AMM of baryons is ignored, upsilon states exhibit larger mass shift as compared to the case where AMM effects are incorporated. A discussion of the mass shifts of upsilon states in the magnetized nuclear matter at T=0 MeV without considering the finite quark mass term is given in the Ref \cite{Amal2_upsilon}. At T=0 MeV, the medium modification of upsilon states is quite similar to that of charmonium states investigated in Ref \cite{Magstrange}. 

The positive mass modification due to the magnetic field decreases as the temperature increases from T=0 MeV to T=90 MeV. Above T=90 MeV, the value of $\chi$ instead drops with an increase in the magnetic field. Hence the magnitude of the negative mass shift also increases with the magnetic field. For $f_s$=0.3 and T=150 MeV, including the effects of AMM, the mass shift of $\Upsilon(4S)$ at $\rho_B$= 4$\rho_0$ is $-243.87$ MeV at eB= $4m_\pi^2$, which increases to $-279.91$ MeV at eB= $8m_\pi^2$. At T=150 MeV, when AMM of baryons is ignored, upsilon states exhibit smaller mass modifications as compared to the case where AMM effects are incorporated. The effects of AMM on the mass shift of upsilon states are observed to be larger at high temperatures.

\section{SUMMARY}
To summarize, we have investigated the mass modifications of open bottom mesons and upsilon states in hot, isospin asymmetric strange hadronic matter in strong magnetic fields using a chiral effective Lagrangian model. To study the magnetized hadronic matter, we have solved the equations of motion of scalar fields as functions of baryon density at various values of temperature for different values of strangeness fraction and magnetic fields. The scalar meson fields ($\sigma$, $\zeta$, and $\delta$) mimic the light quark condensates in the medium, and the dilaton field ($\chi$) simulates the gluon condensates in the medium. From the medium modifications of the scalar mesons fields, number density, and scalar density of the baryons, we obtain the in-medium masses of the open bottom mesons. The charged ${B}^+$ and ${B}^-$ mesons undergo additional positive mass shifts due to the Landau quantization effect in the presence of the magnetic field. The mass shifts of upsilon states are attributed to the medium modification of the dilaton field and the finite quark mass term. The open bottom mesons and upsilon states experience a mass drop in the magnetized nuclear medium compared to their vacuum masses, and the magnitude of this mass drop increases with baryon density. When hyperons are added to the medium, the intensity of this mass drop becomes more significant, and the mass degeneracy of $B_s$ mesons is broken. 

The in-medium masses of open bottom meson and upsilon states are observed to initially increase with a temperature rise and subsequently decrease with a further increase in temperature. The value of this turnaround temperature at which this functional behavior reverses depends upon the strength of the magnetic field, baryon density, and the strangeness fraction. The increase in the magnetic field results in larger in-medium masses of these heavy flavor mesons when the temperature is moderate. However, at higher temperatures, the in-medium masses of all mesons become smaller with an increase in the strength of the magnetic field. The mass shifts of upsilon states as a function of the magnetic field are marginal when AMM effects are neglected. In the present investigation of open bottom mesons and upsilon states, the dominant medium effect is due to the density compared to the magnetic field. These medium modifications should have observable consequences in the ${B}^+$/${B}^0$, ${B}^-$/ $\bar{B^0}$, and ${B_{s}}^0$/ $\bar{{B_{s}}}^0$ ratios in asymmetric heavy-ion collisions in Compressed Baryonic Matter (CBM) experiments at FAIR at the future facility of GSI. The mass shifts of upsilon states can show in the dilepton spectra in these experiments when there is access to higher energies compared to the current planned energy range at CBM.

\section{ACKNOWLEDGEMENTS}
A.J.C.S acknowledges the support towards this work from the Department of Science and Technology, Government of India, via an INSPIRE fellowship [INSPIRE Code IF170745]. A.J.C.S is thankful to Ankit Kumar, Pallabi Parui, and Sourodeep De for fruitful discussions. AM acknowledges financial support from Department of Science and Technology (DST), Government of India [project no.CRG/2018/002226].

\end{document}